\documentclass[journal]{IEEEtran}
\IEEEoverridecommandlockouts
\def\linspread{1}
\def\linspreadalgr{1}
\usepackage{tablefootnote}
\usepackage{cite}
\usepackage{color}
\usepackage{amsfonts}
\usepackage{amssymb}
\usepackage{amsmath}
\usepackage{amsthm}
\usepackage{bm}
\usepackage{algorithm}
\usepackage{algorithmic}
\usepackage{array}
\usepackage{booktabs} 
\usepackage{float} 
\usepackage{multirow}
\usepackage{balance}  
\usepackage{graphicx}
\usepackage{subcaption}  
\usepackage{textcomp}
\usepackage{stfloats}  
\usepackage{url}
\usepackage{verbatim} 
\usepackage{graphicx}
\usepackage{epstopdf}
\usepackage{lettrine}
\usepackage{balance}  
\usepackage[colorlinks=false,
linkcolor=blue,
anchorcolor=blue,
citecolor=blue]{hyperref}
\graphicspath{ {figs/system-model/} {figs/simulation/} {figs/biography/} }
\DeclareMathOperator*{\argmax}{arg\,max}
\DeclareMathOperator*{\argmin}{arg\,min}
\theoremstyle{definition}

\theoremstyle{remark}

\allowdisplaybreaks[3]
\usepackage{afterpage}

\def\BibTeX{{\rm B\kern-.05em{\sc i\kern-.025em b}\kern-.08em
		T\kern-.1667em\lower.7ex\hbox{E}\kern-.125emX}}

\makeatletter

\newcommand{\Rmnum}[1]{\expandafter\@slowromancap\romannumeral #1@}
\makeatother

\newfloat{figtab}{thb}{fgtb}
\makeatletter
\newcommand\figcaption{\def\@captype{figure}\caption}
\newcommand\tabcaption{\def\@captype{table}\caption}
\makeatother

\begin{document}

\title{Near-Field Sparse Bayesian Channel Estimation and Tracking for XL-IRS-Aided Wideband mmWave Systems
}

\author{Xiaokun Tuo, Zijian Chen, Ming-Min Zhao, Changsheng You, and Min-Jian Zhao
\thanks{X. Tuo, Z. Chen,  M. M. Zhao and M. J. Zhao  are with the College of Information Science and Electronic Engineering, Zhejiang University, Hangzhou 310027, China and also with the Zhejiang Provincial Key Laboratory of Multi-Modal Communication Networks and Intelligent Information Processing, Hangzhou 310027, China (e-mail: xktuo@zju.edu.cn; chenzijian@zju.edu.cn; zmmblack@zju.edu.cn; mjzhao@zju.edu.cn). 

C. You is with the Department of Electronic and Electrical Engineering, Southern University of Science and Technology (SUSTech), Shenzhen 518055, China (e-mail: youcs@sustech.edu.cn).
		}}
\maketitle

\begin{abstract}
The rapid development of 6G systems demands advanced technologies to boost network capacity and spectral efficiency, particularly in the context of intelligent reflecting surfaces (IRS)-aided millimeter-wave (mmWave) communications. A key challenge here is obtaining accurate channel state information (CSI), especially with extremely large IRS (XL-IRS), due to near-field propagation, high-dimensional wideband cascaded channels, and the passive nature of the XL-IRS. In addition, most existing CSI acquisition methods fail to leverage the spatio-temporal sparsity inherent in the channel, resulting in suboptimal estimation performance. To address these challenges, we consider an XL-IRS-aided wideband multiple-input multiple-output orthogonal frequency division multiplexing (MIMO-OFDM) system and propose an efficient channel estimation and tracking (CET) algorithm. Specifically, a unified near-field cascaded channel representation model is presented first, and a hierarchical spatio-temporal sparse prior is then constructed to capture two-dimensional (2D) block sparsity in the polar domain, one-dimensional (1D) clustered sparsity in the angle-delay domain, and temporal correlations across different channel estimation frames. Based on these priors, a tensor-based sparse CET (TS-CET) algorithm is proposed that integrates tensor-based orthogonal matching pursuit (OMP) with particle-based variational Bayesian inference (VBI) and message passing. Simulation results demonstrate that the TS-CET framework significantly improves the estimation accuracy and reduces the pilot overhead as compared to existing benchmark methods.

\end{abstract}

\begin{IEEEkeywords}
	Intelligent reflecting surface, near-field communications, channel estimation and tracking, MIMO-OFDM, mmWave, Bayesian inference.
\end{IEEEkeywords}

\vspace{-0.0cm}

\section{Introduction}
\label{sec:Introduction}
\lettrine[lines=2]{A}{chieving} low-cost, energy-efficient 6G networks that also deliver enhanced spectral efficiency, resource utilization, and intelligent sensing has become a central objective of current wireless research \cite{saa20206G}.
To achieve these ambitious goals, researchers are actively exploring various enabling technologies, among which massive multiple-input multiple-output (MIMO) and its more advanced frontier, extremely large-scale MIMO (XL-MIMO), have attracted extensive attention from both academia and industry owing to its potential to significantly enhance network capacity and spectral efficiency \cite{cui2022nearfield}.
Owing to the dramatic increase in the antenna number and the operation at millimeter-wave (mmWave) even terahertz frequency bands, the Rayleigh distance expands considerably, causing both transceivers and scatterers to fall within the near-field region of the XL-MIMO array \cite{Lu2024NF}.
Consequently, electromagnetic propagation transitions from planar to spherical wavefronts, which fundamentally transforms the associated channel modeling and signal processing frameworks \cite{cui2022nearfield,yu2023mixednf}.
Despite their great potential, the large number of antennas and radio-frequency (RF) chains in XL-MIMO systems lead to high hardware cost and energy consumption, posing a major challenge to practical deployment \cite{Lu2024NF}.
To address this issue, extremely large intelligent reflecting surfaces (XL-IRSs) have recently emerged as a promising solution \cite{Wu2021IRS,you20256G}.
By intelligently tuning their reflection coefficients, XL-IRS can generate high-gain passive beams with low cost and power consumption, thereby achieving performance comparable to that of XL-MIMO systems.

Given the importance of channel state information (CSI) acquisition in XL-IRS-aided systems, existing works have conducted in-depth investigations of channel estimation (CE) under both narrowband and wideband settings \cite{Wei2022narrow,Zheng2022CSCE,yang2024nearIRS,Wu2023parametric}. However, as 6G networks continue to demand higher data rates and broader bandwidths, wideband transmission is becoming a fundamental requirement, thereby making wideband channel estimation increasingly critical in XL-IRS-aided communications \cite{saa20206G}.
Motivated by this need, various methods have been proposed to improve channel estimation performance in wideband scenarios. Specifically, a structured canonical polyadic decomposition (CPD)-based channel estimation method has been proposed for IRS-assisted mmWave multiple-input single-output (MISO) orthogonal frequency-division multiplexing (OFDM) systems, enabling unique recovery of the wideband cascaded channel with $\mathcal{O}(L^2)$ training overhead, where $L$ is the number of paths \cite{Zheng2022CSCE}. 
Furthermore, to mitigate the beam-squint effect inherent in wideband XL-IRS-aided mmWave systems, a multi-carrier spherical-domain dictionary was designed, in which the CE task was formulated as a compressed sensing (CS)-based parameter recovery problem solved via atomic correlation matching \cite{yang2024nearIRS}. 
In addition, a polar-domain frequency-dependent CE scheme was developed for terahertz XL-IRS systems, achieving notable gains over conventional approaches \cite{Wu2023parametric}.
Despite these advances, few studies have leveraged the inherent spatial-domain and delay-domain sparsity of near-field cascaded channels in wideband systems to enhance CSI acquisition accuracy. Moreover, in scenarios with slow user mobility, the IRS-user channel exhibits strong temporal correlation, which remains largely underutilized for pilot overhead reduction.

Motivated by the aforementioned considerations, this paper investigates near-field channel estimation and tracking (CET) in XL-IRS-assisted wideband mmWave systems.
We propose a tensor-based sparse CET (TS-CET) algorithm, which models the CET problem in the polar-angle-delay domain to achieve accurate CSI acquisition with low pilot overhead.
The main contributions of this paper are summarized as follows.
\begin{itemize}
    \item A unified near-field cascaded channel model is developed which jointly captures spherical-wave propagation, Doppler-induced phase variations, and the temporal evolution of the channels. This modeling provides a solid foundation for the design of subsequent TS-CET algorithm.

    \item A hierarchical spatio-temporal sparse prior is formulated to capture the structural characteristics of the hybrid near- and far-field wideband cascaded channel, which jointly models the two-dimensional (2D) block sparsity in the polar domain, the one-dimensional (1D) clustered sparsity in the angular and delay domains, and the inter-frame temporal correlation, thereby fully integrating the spatial, frequency, and temporal structural sparsities within a unified framework to enhance estimation accuracy.

    \item A two-stage CET scheme is proposed.
    Specifically, in the CE phase, a tensor-based orthogonal matching pursuit (OMP) algorithm is employed for low-complexity coarse estimation, followed by particle-based variational Bayesian inference (VBI) with message passing for refinement.
    In the channel tracking (CT) phase, spatio-temporal priors and historical channel information are exploited to approximate the joint posterior distribution of the the sparse cascaded channel gains and non-ideal parameters for dynamic channel tracking. 

    \item Simulation results demonstrate that the proposed TS-CET framework outperforms the existing baseline methods, including expectation-maximization OMP (EM-OMP) and expectation-maximization fast sparse Bayesian learning (EM-FSBL), achieving higher estimation and tracking accuracy as well as improved pilot efficiency.
\end{itemize}
The rest of this paper is organized as follows. 
Section~\ref{sec:System_Model} introduces the system model and the two-phase transmission protocol, which includes the CE and CT phases.
Section~\ref{sec:Hierarchical_Prior_Model} presents the hierarchical spatio-temporal sparse prior model and the design of active and passive beamforming for the CT phase.
Section~\ref{sec:TS-CET_Algorithm} elaborates on the proposed TS-CET algorithm for CE and CT.
Section~\ref{sec:simulation} presents numerical results to validate the effectiveness of the proposed framework.
Finally, Section~\ref{sec:conclusion} concludes the paper.
 \begin{figure}[htbp]
 	\setlength{\abovecaptionskip}{-0.0cm}
 	\centering
 	{\includegraphics[width=0.45\textwidth]{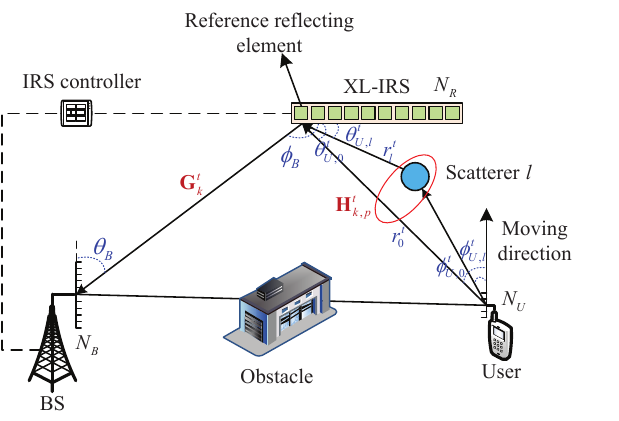}}
 	\caption{System model of the considered XL-IRS-aided wideband mmWave system.}
 	\label{pic:sys_mod}
 	\vspace{-0.0cm}
 \end{figure}
\par \textit{Notations:} Scalars, vectors and matrices are respectively denoted by lower/upper case, boldface lowercase and boldface uppercase letters. The  transpose, conjugate and conjugate transpose of a general vector $\mathbf{z}$ are denoted as $\mathbf{z}^T$, $\mathbf{z}^*$ and $\mathbf{z}^H$, respectively. $\text{diag}(\mathbf{z})$ yields a diagonal matrix with $\mathbf{z}$ as its diagonal elements.    
$\delta(\cdot)$ denotes the Dirac delta function. The indicator function $\mathbb{I}(x_1,\cdots,x_N)$ is defined such that $\mathbb{I}(x_1,\cdots,x_N) = 1$, if $x_n = 1$, $\forall n = 1,\cdots, N$; otherwise, $\mathbb{I}(x_1,\cdots,x_N) = 0$.   $\circ$, $\diamond$, $\otimes$ and $\odot$  denote the tensor outer product, Hadamard product, Kronecker product and Khatri-Rao product, respectively. For an $N$-order tensor $\bm{\mathcal{X}} \in \mathbb{C}^{ I_1\times I_2 \times \cdots \times I_N }$, the mode-$n$ unfolding matrix $\mathbf{X}_{(n)} \in \mathbb{C}^{ I_n \times \prod_{m\neq n} I_m }$ can be written as $[\mathbf{X}_{(n)}]_{i_n, j } = [\bm{\mathcal{X}}]_{i_1,\cdots,i_N}$ with $j = 1 + \sum_{k=1, k\neq n}^N (i_k -1)J_k$ and $J_k = \prod_{m=1, m\neq n}^{k-1} I_m$, where $i_n = 1, \cdots, I_n$ and $n = 1, \cdots, N$. The Tucker decomposition of $\bm{\mathcal{X}}$ is given by \cite{kolda2009tensor}
\begin{equation}
	\label{eq:Tucker decom}
	\bm{\mathcal{X}} = 	\bm{\mathcal{Z}} \times_1 \mathbf{D}_1 \times_2 \mathbf{D}_2  \times_3 \cdots \times_N \mathbf{D}_N,
\end{equation}
where $\bm{\mathcal{Z}} \in \mathbb{C}^{K_1\times \cdots \times K_N}$ is the core tensor and $\mathbf{D}_n$ is an $I_n \times K_n$ matrix. The operation $\times_n$ denotes the $i$-mode product for a tensor. Specifically, for the case of $\bm{\mathcal{Y}} = \bm{\mathcal{X}} \times_n \mathbf{D}$, its  equivalent operation can be written as $\mathbf{Y}_{(n)} = \mathbf{D}\mathbf{X}_{(n)}$. Based on \eqref{eq:Tucker decom}, the vectorized form of $	\bm{\mathcal{X}} $ can be expressed as
\begin{equation}
	\label{eq:vec of tensor}
	\mathbf{x} = \text{vec}(\bm{\mathcal{X}} ) = \Big( \bigotimes_{n}\mathbf{D}_n \Big) \mathbf{z},
\end{equation}
where $\mathbf{z}$ is the vectorized form of $\bm{\mathcal{Z}}$ and  $\bigotimes_{n=1}^N\mathbf{D}_n = \mathbf{D}_N \otimes \mathbf{D}_{N-1} \otimes \cdots \otimes\mathbf{D}_{1}$  denotes the sequential Kronecker product.

\vspace{-0.0cm}
\section{System Model and Proposed Transmission Protocol}
\label{sec:System_Model}
\subsection{System Model}
\par As shown in Fig. \ref{pic:sys_mod}, we consider a time division
duplexing (TDD) wideband mmWave system, where an XL-IRS composed of
$N_R$ reflecting elements is deployed near a user\footnote{The framework can be extended to the multiuser scenario by allowing each user to transmit orthogonal pilot sequences, thereby enabling the separation and estimation of the cascaded channels for each user.} to enhance the communication between the user and the BS. The BS is equipped with $N_B$ antennas and $R_B$ RF chains ($R_B \ll N_B$), and the user is equipped with $N_U$ antennas and a single RF chain. The BS antennas, user antennas and XL-IRS reflecting elements are arranged as uniform linear arrays (ULAs) with the antenna/element interval
being half of the carrier wavelength $\lambda_c$. To address the frequency-selective fading of the wideband channels, the considered system adopts $K$-subcarrier OFDM modulation and the frequency of the $k$-th subcarrier is $f_k = \frac{f_s}{K}(k-1-\frac{K-1}{2})$, $\forall k \in \mathcal{K} \triangleq \{1, \cdots, K\}$, with $f_s$ being the bandwidth.  
We assume in this paper that the user moves at a low speed $v$ leading to a maximum Doppler frequency offset (DFO) $f_d = \frac{v}{\lambda_{c}}$.
Besides, we further assume that the direct link between the BS and the user is blocked by certain obstacles (e.g., walls and buildings). 

\subsection{Transmission Protocol}
\par
\begin{figure}[tbp]
	\setlength{\abovecaptionskip}{-0.0cm}
	\centering
	{\includegraphics[width=0.49\textwidth]{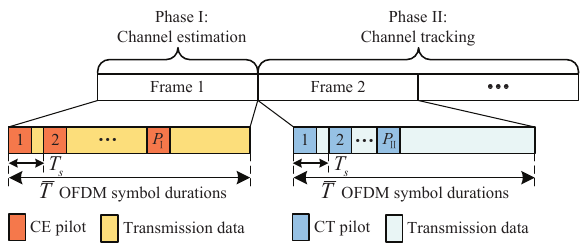}}
	\caption{Transmission protocol.}
	\label{pic:frame_structure}
	\vspace{-0.0cm}
\end{figure}

 Due to user mobility, the cascaded BS-IRS-user channel is time-varying and exhibits a substantial level of temporal correlation. Therefore, our goal is to first estimate this cascaded channel using a relatively large number of pilots due to the lack of prior channel information. Then, we aim to track this channel with a significantly reduced number of pilots by leveraging both the estimation results and the temporal correlation of the cascaded channel. To this end, we propose a specific transmission protocol comprising two phases, as depicted in Fig. \ref{pic:frame_structure}. Specifically, the considered timeline is divided into frames, with each frame consisting of $\bar{T}$ OFDM symbols. It is assumed that the channel parameters (as will be specified later) remain unchanged within each frame. Note that $\text{phase I}$ comprises only the initial frame (i.e., frame 1), while the subsequent frames (i.e., frame $t$, $t = 2,3, \cdots$) are allocated to $\text{phase II}$. In phase I, the user transmits $P_{\text{I}}$ OFDM pilots at an interval of $T_s$, which results in a larger phase rotation of the received signal at the BS due to the DFO compared to squeezing all the pilots at the beginning of phase I. Therefore, employing this comb-structured pilot pattern can enhance the DFO estimation performance. Based on the overall observations received at the BS, the cascaded BS-IRS-user channel can be estimated (the details will be shown in Section ~\ref{sec:TS-CET_Algorithm}). Then, in frame $t$ ($\forall t \geq 2$) of $\text{phase II}$, $P_{\text{II}}$ ($P_{\text{II}} < P_{\text{I}}$) OFDM pilots are transmitted following the same procedure as in $\text{phase I}$. In this phase, considering the temporal correlation of the cascaded channel and by treating the estimation result obtained from the $(t-1)$-th frame as the prior information in the $t$-th frame, we propose to track the cascaded channel with reduced pilot overhead (as will be elaborated in Section ~\ref{sec:TS-CET_Algorithm}). Besides, in the remaining duration of each frame, uplink data transmission is carried out between the BS and the user.

\subsection{Channel Model}
In traditional far-field channel modeling, the array response vector (ARV) for an $N$-element ULA under the planar-wave assumption can be expressed as
\begin{equation}
	\mathbf{a}^{far}(\vartheta) = [1, e^{-j\frac{2\pi}{\lambda_c} d \vartheta}, \cdots,  e^{-j\frac{2\pi}{\lambda_c} (N-1) d \vartheta} ]^T,
\end{equation}
where $d$ denotes the element interval and $\vartheta = \cos\theta$ is the spatial angle with $\theta$ being the physical angle-of-arrival (AoA) or angle-of-departure (AoD). While for an XL-IRS operating in the mmWave band, the near-field region characterized by the Rayleigh distance $Z = \frac{2D^2}{\lambda_c}$ with $D$ representing the array aperture is substantially extended due to the large number of reflecting elements and the small carrier wavelength. Hence, objects present in the environment, such as communication scatterers and users, are more likely to be located within the near-field region. In such scenarios, the far-field ARV fails to characterize  the propagation characteristics of the near-field spherical electromagnetic wave \cite{cui2022nearfield}. To address this issue, we adopt the near-field ARV under the spherical-wave assumption, which is given by \cite{yu2023mixednf} 
\begin{equation}
	\mathbf{a}^{near}(\varphi,r) = [e^{-j\frac{2\pi}{\lambda_c} (r^{(1)} - r) }, \cdots, e^{-j\frac{2\pi}{\lambda_c} (r^{(N)} - r) } ]^T,
\end{equation}
where $\varphi$ and $r$ denote the spatial angle and the distance from the reference element to the scatterer/user, respectively; $r^{(n)}, \, \forall n \in \{1, \cdots, N\}$, denotes the distance from the $n$-th element to the scatterer/user and can be expressed as
\begin{align}
	r^{(n)} &= \sqrt{ (r \sqrt{1-\varphi^2} )^2 + (r\varphi - (n-1)d )^2   } \nonumber \\
	& \overset{(a)}{\approx} r - d(n-1)\varphi + \frac{d^2(n-1)^2}{2}\frac{1-\varphi^2}{r},
\end{align}
where $(a)$ is due to the Fresnel approximation \cite{Selvan2017fraun}, i.e., $\sqrt{1+x} = 1 + \frac{1}{2}x - \frac{1}{8}x^2 + \mathcal{O}(x^2)$.

\par  By taking the DFO into consideration, the $k$-th subcarrier IRS-user channel in the $p$-th pilot duration of frame $t$ ($t = 1,2,\cdots$) can be expressed as\footnote{Since the number of antennas at the user is typically small due to the restrictions in power consumption and hardware cost, we adopt a far-field assumption at the user side as in \cite{yang2024nearIRS}.}
\begin{equation}
\label{IRS-user channel}
\mathbf{H}^{(t)}_{k,p} \!\!=\!\! \sum_{l=0}^{L^{(t)}\!-\!1} \!\! \alpha_l^{(t)} e^{-j2\pi f_k\tau_{U,l}^{(t)}}e^{j\psi_{p,l}^{(t)}}\mathbf{a}_{R}(\vartheta_{U,l}^{(t)}, r_{U,l}^{(t)}) \mathbf{a}_{U}^H(\varphi_{U,l}^{(t)}), 
\end{equation}
where $p \in \mathcal{P} \triangleq \{ 1, \cdots, P \}$ with  $P = P_{\text{I}}$ for phase I and $ P = P_{\text{II}} $ for phase II;
$L^{(t)}$ denotes the total number of propagation paths; 
$\alpha_l^{(t)}$ and $\tau_{U,l}^{(t)}$  denote the complex-valued path gain and path delay associated with the $l$-th path, respectively; 
$\psi_{p,l}^{(t)} = 2\pi f_d^{(t)} (p-1) T_s \varphi_{U,l}^{(t)}$ denotes the $l$-th path phase rotation caused by the DFO $f_d^{(t)} \varphi_{U,l}^{(t)}$ with $f_d^{(t)}$ being the maximum DFO and $\varphi_{U,l}^{(t)} = \cos\phi_{U,l}^{(t)}$ being the $l$-th path spatial angle associated with its physical AoD $\phi_{U,l}^{(t)}$; 
$\mathbf{a}_{R}(\vartheta_{U,l}^{(t)}, r_{U,l}^{(t)})$ and $\mathbf{a}_{U}(\varphi_{U,l}^{(t)})$ denote the near-field ARV and far-field ARV at the XL-IRS and user, respectively;
$\vartheta_{U,l}^{(t)} = \cos \theta_{U,l}^{(t)}$ denotes the $l\text{-th}$ spatial angle  associated with its physical AoA $\theta_{U,l}^{(t)}$;  
$r_{U,l}^{(t)}$ denotes the distance from the user ($l=0$) or the $l\text{-th}$ ($l\neq0$) scatterer to the XL-IRS reference element. Note that although the channel parameters $L^{(t)}$, $\alpha_l^{(t)}$, $\tau_{U,l}^{(t)}$, $f_d^{(t)}$, $\vartheta_{U,l}^{(t)}$, $\varphi_{U,l}^{(t)}$, and $r_{U,l}^{(t)}$ remain constant within the current frame $t$, the IRS-user channel $\mathbf{H}^{(t)}_{k,p}$ varies across different OFDM pilots due to the Doppler term $e^{j\psi_{p,l}^{(t)}}$. 
\par On the other hand, as the XL-IRS is deployed near the user to enhance its transmission ability, resulting in a considerable distance between the BS and XL-IRS, the $k$-th subcarrier BS-IRS channel in the $t$-th frame is modeled as a far-field line-of-sight (LoS) channel\cite{pan2023irsloc}, i.e., 
\begin{equation}
	\label{BS-IRS channel}
	\mathbf{G}_k^{(t)} = \beta^{(t)} e^{-j 2\pi f_k \tau_{B}} \mathbf{a}_{B}(\vartheta_{B}) \mathbf{a}_{R}^H(\varphi_{B}),
\end{equation}
where $\beta^{(t)} $ and $\tau_{B}$ denote the corresponding complex path gain and path delay, respectively;
$\mathbf{a}_{B}(\vartheta_{B})$ and $\mathbf{a}_{R}(\varphi_{B})$ denote the far-field ARVs at the BS and XL-IRS, respectively;
$\vartheta_{B} = \cos \theta_{B}$ and $\varphi_{B} = \cos \phi_{B} $ denote the spatial angles at the BS and XL-IRS, respectively, with $\theta_{B}$ and $\phi_{B} $ being the corresponding physical AoA and AoD. In this paper, we assume that the path delay $\tau_B$ and the spatial angles $\vartheta_{B}$ and $\varphi_{B}$, derived from the geometric relations between the BS and XL-IRS, remain unchanged across all frames due to the fixed positions of the BS and XL-IRS. 

\par In the $p$-th pilot duration of frame $t$, let $x_{k,p} \in \mathbb{C}$  denote the uplink pilot transmitted on the $k$-th subcarrier, and let $\mathbf{f}_{p}^{(t)}  \in \mathbb{C}^{N_U \times 1}$ and  $\mathbf{W}_p^{(t)} \in \mathbb{C}^{N_B \times R_B}$  denote the active beamforming vector and matrix at the user and BS, respectively. Then, the BS received signal on the $k$-th subcarrier, denoted by $\mathbf{y}_{k,p}^{(t)}$, is given by 
\vspace{-0.0cm}
\begin{align}
	\label{eq:BS received signal}
	\mathbf{y}_{k,p}^{(t)} &= \sqrt{p_T}\big(\mathbf{W}_{p}^{(t)}\big)^{\!H} \mathbf{G}_k^{(t)} \boldsymbol{\Theta}_p^{(t)} \mathbf{H}_{k,p}^{(t)} \mathbf{f}_p^{(t)} x_{k,p} + \bar{\mathbf{n}}_{k,p}^{(t)} \nonumber \\
	& = \sqrt{p_T}x_{k,p}\!\Big(\! \big(\mathbf{f}_p^{(t)}\big)^{\!\!T}\! \!\!\otimes\!\! \big(\!\mathbf{W}_{p}^{(t)}\big)^{\!\!H} \Big)\! \Big(\! \big(\mathbf{H}_{k,p}^{(t)}\big)^{\!\!T}\!\! \!\odot\! \mathbf{G}_k^{(t)}   \!  \Big)\! \mathbf{v}_p^{(t)} \!\!+\! \bar{\mathbf{n}}_{k,p}^{(t)}  \nonumber \\
	& =\sqrt{p_T}x_{k,p}\mathbf{X}_p^{(t)} \mathbf{R}_{k,p}^{(t)} \mathbf{v}_p^{(t)} + \bar{\mathbf{n}}_{k,p}^{(t)},  
\end{align}
where $p_T$ denotes the transmit power at the user; $\mathbf{v}_p^{(t)} \triangleq [e^{j\iota_{1,p}^{(t)}}, \cdots, e^{j\iota_{N_R,p}^{(t)}}]^T $ denotes the XL-IRS reflection vector (also referred to as the passive beamforming vector) with $\iota_{n,p}^{(t)}$, $\forall n \in \mathcal{N}_R \!\triangleq\! \{  1, \cdots, N_R  \} $, being the phase shift of the $n$-th reflecting element;
$\boldsymbol{\Theta}_p^{(t)} \triangleq \text{diag}(\mathbf{v}_p^{(t)} )$ denotes the XL-IRS reflection matrix;  
$\mathbf{R}_{k,p}^{(t)} \triangleq \big(\mathbf{H}_{k,p}^{(t)}\big)^{\!T}\!\! \odot\! \mathbf{G}_k^{(t)}$  denotes the cascaded channel;
$\mathbf{X}_p^{(t)}$ is defined as $\mathbf{X}_p^{(t)} \triangleq \big(\mathbf{f}_p^{(t)}\big)^{\!T}\! \otimes\! \big(\mathbf{W}_{p}^{(t)}\big)^{\!H}$;
$\bar{\mathbf{n}}_{k,p}^{(t)} \triangleq \big(\mathbf{W}_{p}^{(t)}\big)^{\!H} \mathbf{n}_{k,p}^{(t)}$, and $\mathbf{n}_{k,p}^{(t)} \sim \mathcal{CN}(\mathbf{0}, \sigma^2\mathbf{I})$ denotes the additive white Gaussian noise (AWGN). Recalling \eqref{IRS-user channel} and \eqref{BS-IRS channel}, the cascaded channel $\mathbf{R}_{k,p}^{(t)}$ can be further expressed as
\vspace{-0.0cm}
\begin{align}
	\label{eq:cas_chan}
	\mathbf{R}_{k,p}^{(t)} & = \sum_{l=0}^{L^{\!(t)}\!-\!1}\alpha_l^{(t)} \beta^{(t)} e^{-j 2\pi f_k (\tau_{U,l}^{(t)} + \tau_{B})} e^{j \psi_{p,l}^{(t)}} \nonumber \\
	& \quad   \; \! \times \! \big( \mathbf{a}_{U}^*(\varphi_{U,l}^{(t)}) \mathbf{a}_{R}^T(\vartheta_{U,l}^{(t)}, r_{U,l}^{(t)})  \big) \!\odot\! \big(  \mathbf{a}_{B}(\vartheta_{B}) \mathbf{a}_{R}^H( \varphi_{B} )  \big)  \nonumber \\
	& \overset{(a)}{=}\! \sum_{l=0}^{L^{\!(t)}\!-\!1} \! \gamma_l^{(t)} e^{-j 2\pi f_k \bar{\tau}_l^{(t)}}  e^{j \psi_{p,l}^{(t)}}  \nonumber \\
	& \quad  \; \times  \!  \big( \mathbf{a}_{U}^*(\varphi_{U,l}^{(t)}) \!\otimes\! \mathbf{a}_{B}(\vartheta_{B})  \big)  \big(  \mathbf{a}_{R}^T(\vartheta_{U,l}^{(t)}, r_{U,l}^{(t)}) \! \odot \! \mathbf{a}_{R}^H( \varphi_{B} )   \big) \nonumber \\
	& \;=\sum_{l=0}^{L^{\!(t)}\!-\!1} \! \gamma_l^{(t)} e^{-j 2\pi f_k \bar{\tau}_l^{(t)}} \tilde{\mathbf{a}}_{BU,p}(\varphi_{U,l}^{(t)}) \mathbf{a}_{R}^T(\bar{\vartheta}_{l}^{(t)}, \bar{r}_{l}^{(t)}),
\end{align}
where $ \gamma_{l}^{(t)} \!\triangleq\! \alpha_l^{(t)} \beta^{(t)}$ and $\bar{\tau}_l^{(t)} \!\triangleq\! \tau_{U,l}^{(t)} + \tau_{B}$ denote the cascaded path gain and delay, respectively;
$(a)$ follows from $\mathbf{A}\mathbf{B} \odot \mathbf{C}\mathbf{D} = (\mathbf{A} \otimes \mathbf{C})(\mathbf{B} \odot \mathbf{D})$;
$\tilde{\mathbf{a}}_{BU,p}(\varphi_{U,l}^{(t)}) \triangleq \mathbf{a}_{BU}(\varphi_{U,l}^{(t)})  e^{j \psi_{p,l}^{(t)}}$ and  $\mathbf{a}_{BU}(\varphi_{U,l}^{(t)}) \triangleq \mathbf{a}_{U}^*(\varphi_{U,l}^{(t)}) \otimes \mathbf{a}_{B}(\vartheta_{B}) $;
$\mathbf{a}_{R}^T(\bar{\vartheta}_{l}^{(t)}, \bar{r}_{l}^{(t)}) \triangleq  \mathbf{a}_{R}^T(\vartheta_{U,l}^{(t)}, r_{U,l}^{(t)})  \odot \mathbf{a}_{R}^H( \varphi_{B} ) $ with $\bar{\vartheta}_{l}^{(t)} \triangleq  \vartheta_{U,l}^{(t)} + \varphi_{B}$ and $\bar{r}_{l}^{(t)} \triangleq \frac{1-(\bar{\vartheta}_{l}^{(t)})^2}{1-(\vartheta_{U,l}^{(t)})^2}r_{U,l}^{(t)}$ being the effective spatial angle and distance, respectively. Note that in the above derivation of $\mathbf{a}_{R}^T(\bar{\vartheta}_{l}^{(t)}, \bar{r}_{l}^{(t)})$, we use the fact that the phase of the $n$-th element of $\mathbf{a}_{R}^T(\vartheta_{U,l}^{(t)}, r_{U,l}^{(t)})  \odot \mathbf{a}_{R}^H( \varphi_{B} )$, which can be expressed as 
\vspace{-0.0cm}
\begin{align}
	& \big(r_{U,l}^{{(t)}}\big)^{(n)}-r_{U,l}^{(t)} - d(n-1)\varphi_{B} \nonumber \\
	&=-d(n-1)\bar{\vartheta}_{l}^{(t)} + \frac{d^2(n-1)^2}{2} \frac{1-(\bar{\vartheta}_{l}^{(t)})^2}{\bar{r}_{l}^{(t)}},
\end{align}
is exactly the phase of the $n$-th element of $\mathbf{a}_{R}^T(\bar{\vartheta}_{l}^{(t)}, \bar{r}_{l}^{(t)})$.  By defining $\mathbf{a}_f(\tau)  \triangleq  [ e^{-j 2\pi f_1\tau},  \!\cdots\!,   e^{-j 2\pi f_K\tau}   ]^T$, the cascaded channel for all the $K$ subcarriers can be expressed as a three-order tensor $\bm{\mathcal{R}}_p^{(t)} \in \mathbb{C}^{N_{BU} \times N_R \times K}$, i.e., 
\begin{equation}
	\label{eq:four-order cas_chan}
	\bm{\mathcal{R}}_p^{(t)} \!=\!\!\! \sum_{l=0}^{L^{(t)}-1} \! \gamma_{l}^{(t)}  \tilde{\mathbf{a}}_{BU,p}(\varphi_{U,l}^{(t)}) \circ  \mathbf{a}_{R}(\bar{\vartheta}_{l}^{(t)}, \bar{r}_{l}^{(t)}) \circ  \mathbf{a}_{f}(\bar{\tau}_{l}^{(t)}),
\end{equation}
where $N_{BU} \triangleq N_B N_U$ denotes the dimension of $\tilde{\mathbf{a}}_{BU,p}(\varphi_{U,l}^{(t)})$.
It is noteworthy that the correlation between  $\mathbf{R}_{k,p}^{(t)}$ and $	\bm{\mathcal{R}}_p^{(t)} $ is characterized by 
$\mathbf{R}_{k,p}^{(t)} = \big(\big[ \bm{\mathcal{R}}_p^{(t)}  \big]_{:,:,k}\big)_{(1)}$.
Based on \eqref{eq:BS received signal} and \eqref{eq:cas_chan}, the BS received signal for all the $K$ subcarriers during the $p$-th pilot can also be expressed as a three-order tensor $\bm{\mathcal{Y}}_p^{(t)} \in \mathbb{C}^{R_B  \times 1 \times \bar{K}  }$, i.e., 
\begin{align}
	\label{eq:four-order received signal}
	\bm{\mathcal{Y}}_p^{(t)} \!\!=\!\!\! \sum_{l=0}^{L^{(t)}-1} \! & \gamma_{l}^{(t)}  \big(   \mathbf{X}_{p}^{(t)}\tilde{\mathbf{a}}_{BU,p}(\varphi_{U,l}^{(t)}) \big) \circ  \big( (\mathbf{v}_{p}^{(t)})^T  \mathbf{a}_{R}(\bar{\vartheta}_{l}^{(t)}, \bar{r}_{l}^{(t)}) \big) \nonumber \\
	& \circ   \big( \sqrt{p_T} \text{diag}(\mathbf{x}_p)\mathbf{S} \mathbf{a}_{f}(\bar{\tau}_{l}^{(t)}) \big) + \bar{\bm{\mathcal{N}}}_{p}^{(t)},
\end{align}
where $\bar{K}$ subcarriers are utilized for CET and $\mathbf{S} \in \{0,1\}^{\bar{K} \times K}$ is a selection matrix depending on the frequency-domain pilot pattern, e.g., $[\mathbf{S}]_{n_1, n_2} = 1$ for $n_2 = 4n_1-1$, $\forall n_1 \in \bar{\mathcal{K}} \triangleq \{  1, \cdots, \bar{K} \}$, resulting in a comb-4 structured pilot pattern in the frequency domain;
$\mathbf{x}_p = [x_{k_1,p}, \cdots, x_{k_{\bar{K}},p}]^T$ denotes the frequency-domain pilot vector transmitted by the user, with $k_{\bar{k}}$, $\forall \bar{k} \in \bar{\mathcal{K}}$, being the indices of the selected subcarriers, and is obtained from the Zadoff-Chu (ZC) sequence \cite{3GPPzc};
$\bar{\bm{\mathcal{N}}}_{p}^{(t)} \in \mathbb{C}^{R_B  \times 1 \times \bar{K}} $ denotes the AWGN tensor with $(\bar{\bm{\mathcal{N}}}_{p}^{(t)})_{(1)} = [\bar{\mathbf{n}}_{k_1,p}^{(t)}, \cdots, \bar{\mathbf{n}}_{k_{\bar{K}},p}^{(t)}]$.

In summary, spherical wavefronts and Doppler shifts make the XL-IRS-aided wideband cascaded channel markedly different from conventional far-field models, and its high dimensionality, passive reflection, and parameter coupling further pose notable estimation challenges. To address these, we propose to exploit the inherent spatial structure and temporal sparsity by designing an efficient CET algorithm.

\section{Hierarchical Spatio-temporal Sparse Prior Model}
\label{sec:Hierarchical_Prior_Model}
In this section, we develop a Bayesian compressive sensing framework to enable efficient CET.
We first exploit the sparse structure of the cascaded channel $\bm{\mathcal{R}}_p^{(t)}$ across multiple transform domains, including the polar domain at the XL-IRS, the angle domain at the user, and the delay domain. 
For the CE phase, we propose a hierarchical probabilistic model to capture the spatially structured sparsity.
For the subsequent CT phase ($t \ge 2$), we introduce a temporally dynamic probabilistic model to describe the evolution of channel parameters, enabling efficient tracking across frames. Finally, we design both active and passive beamforming strategies. 
The active beams align with known directions, while the passive beams are omnidirectional during the CE phase and trade off enhancing known directions and exploring new ones during the CT phase, thereby improving the overall CET performance across consecutive frames.

\subsection{Polar-Angle-Delay Domain Sparse Representation of the Cascaded Channel}
To obtain the sparse representation of $\bm{\mathcal{R}}_p^{(t)}$ in the polar-angle-delay domain, we first introduce a uniform angle grid $\bm{\varphi} \triangleq [ \varphi_1, \cdots, \varphi_{N_U} ]^T$ consisting of $N_U$ grid points for the angle domain, and a uniform delay grid $\bm{\tau} \triangleq [ \tau_1, \cdots, \tau_{N_f} ]^T$ comprising $N_f$ grid points for the delay domain, where  $\varphi_u$ ($\forall u \in \mathcal{N}_U \triangleq \{1, \cdots, N_U \}$) and $\tau_k$ ($\forall k \in \mathcal{N}_f \triangleq \{1, \cdots, N_f \}$) are given by $\varphi_{u} =  \frac{2}{N_U}\Big( u -  \frac{N_U+1}{2}  \Big)$, $\tau_k = \frac{1}{f_s} \left( k - \frac{1}{2}\right)$.

We assume that $N_f$ is sufficiently large so that all the cascaded path delays are smaller than $\tau_{N_f}$, i.e., $\bar{\tau}_l^{(t)} < \tau_{N_f}$, $\forall t,l$. Next, we introduce a non-uniform position grid $ [\bar{\bm{\vartheta}}, \bar{\mathbf{r}}] \triangleq [\bar{\vartheta}_1, \bar{r}_1; \cdots ; \bar{\vartheta}_{\bar{N}_R}, \bar{r}_{\bar{N}_R}]$ consisting of $\bar{N}_R$ grid points for the polar domain, where $\bar{N}_R = N_R S$ with $N_R$ being the  number of angle grid points and $S$ being the number of distance grid points;
$\bar{\vartheta}_m$ and $\bar{r}_m$ ($\forall m \in \bar{\mathcal{N}}_R \triangleq \{1, \cdots, \bar{N}_R \}$) denote the spatial angle and distance of the $m$-th position grid, respectively, which can be obtained based on the polar-domain sampling method \cite{cui2022nearfield}, i.e., 	
\vspace{-0.15cm}
\begin{equation}
\begin{aligned}
	\bar{\vartheta}_m \!\!=\!\! \frac{2}{N_R}\Big( \Big\lfloor \!\frac{m\!-\!1}{S}\! \Big\rfloor \!\!-\! \frac{N_R\!-\!1}{2}   \Big), \; \bar{r}_m \!\!=\! Z_{\Delta}\frac{1\!-\!\bar{\vartheta}_m^2}{\text{mod}(m\!-\!1,\!S)},
\end{aligned}
\vspace{-0.1cm}
\end{equation}
with $Z_{\Delta}$ being a threshold distance. 

\par In practice, since the spatial angle $\varphi_{U,l}^{(t)}$,  cascaded path delay $\bar{\tau}_l^{(t)}$ and  scatterer/user position $[\bar{\vartheta}_l^{(t)}, \bar{r}_l^{(t)}]$ are continuous, their true values usually  do not  align precisely with the above discrete grid points. To achieve super-resolution estimation of these channel parameters, we introduce  off-grid vectors for the predefined grids.
Specifically, let $u_{l}^{\!(t)} \!=\! \argmin_u | \varphi_{U,l}^{(t)} \!-\! \varphi_u |$,  $k_{l}^{\!(t)} = \argmin_k | \bar{\tau}_{l}^{\!(t)} \!\!\!-\! \tau_k |$ and $m_{l}^{\!(t)} \!\!=\! \argmin_m \! D( [\bar{\vartheta}_l^{(t)}\!\!, \bar{r}_l^{(t)}], [\bar{\vartheta}_m, \bar{r}_m])$, $\forall l \!\in\! \mathcal{L}^{(t)} \!\triangleq\! \{0, \!\cdots\!, L^{(t)}\!-\!1\}$, denote the indices of the predefined grids nearest to $\varphi_{U,l}^{(t)}$, $\bar{\tau}_{l}^{(t)}$ and $[\bar{\vartheta}_l^{(t)}, \bar{r}_l^{(t)}]$, respectively, where $D( [\vartheta_1, r_1], [\vartheta_2, r_2]) = \sqrt{ r_1^2 + r_2^2 - 2r_1r_2 \cos(\theta_2 - \theta_1 )  }$ calculates the distance between polar-domain points $(\vartheta_1, r_1)$ and $(\vartheta_2, r_2)$, with $\theta_1 = \arccos\vartheta_1$ and  $\theta_2 = \arccos\vartheta_2$. Then, the off-grid vectors are defined as $\Delta \bm{\varphi}^{\!(t)} \triangleq [\Delta \varphi_1^{\!(t)}, \cdots, \Delta \varphi_{N_U}^{\!(t)}]^T$, $\Delta \bm{\tau}^{\!(t)} \!\triangleq\! [\Delta \tau_1^{\!(t)}, \!\cdots\!, \Delta \tau_{N_f}^{\!(t)}]^T$, $\Delta \bar{\bm{\vartheta}}^{(t)} \!\triangleq\! [\Delta \bar{\vartheta}_1^{(t)}, \!\cdots\!, \Delta \bar{\vartheta}_{\bar{N}_R}^{(t)}]^T$, and $\Delta \bar{\mathbf{r}}^{(t)} \!\triangleq\! [\Delta \bar{r}_1^{(t)}, \cdots, \Delta \bar{r}_{\bar{N}_R}^{(t)}]^T$, where $\Delta \varphi_u^{(t)}$ ($\forall u \in \mathcal{N}_U$), $\Delta \tau_k^{(t)}$ ($\forall k \in \mathcal{N}_f$), $\Delta \bar{\vartheta}_m^{(t)}$ and $\Delta \bar{r}_m^{(t)}$ ($\forall m \in \bar{\mathcal{N}}_R$) are given by
{\setlength\abovedisplayskip{2pt}
	\setlength\belowdisplayskip{2pt}
\begin{align}
	&\Delta \!\varphi_u^{\!(t)} \!\!\!= \!\! \left\{
	\begin{aligned}
		\!& \varphi_{U,l}^{\!(t)} \!-\! \varphi_{u}, \; \text{if}\; u \!=\! u_{l},  \\ 
		\!& 0,  \; \text{if}\; u \!\neq\! u_{l},
	\end{aligned}
	\right.  
	\Delta \!\tau_k^{\!(t)} \!\!\!=\!\! \left\{
	\begin{aligned}
		\!&\bar{\tau}_{l}^{\!(t)} \!-\! \tau_k, \; \text{if}\; k \!=\! k_{l},  \\ 
		\!& 0,  \; \text{if}\; k \!\neq\! k_{l},
	\end{aligned}
	\right. \\
	&\Delta \!\bar{\vartheta}_m^{(t)}\!\! =\!\! \left\{
	\begin{aligned}
		\!& \bar{\vartheta}_l^{\!(t)} \!\!\!-\!\! \bar{\vartheta}_{\!m}, \; \text{if}\; m \!\!=\!\! m_{l},   \\ 
		\!& 0,  \; \text{if}\; m \!\neq\! m_{l},
	\end{aligned}
	\right.  
	\Delta\! \bar{r}_m^{(t)} \!\!=\!\! \left\{
	\begin{aligned}
		\!&\bar{r}_{l}^{\!(t)} \!\!\!-\!\! \bar{r}_{\!m}, \; \text{if}\; m \!\!=\!\! m_{l},   \\ 
		\!& 0,  \; \text{if}\; m \!\neq\! m_{l},
	\end{aligned}
	\right. 	
\end{align}}%
$ \forall l \in \mathcal{L}^{(t)}$. 

\par Based on the above definitions, the sparse dictionaries for the three modes of $\bm{\mathcal{R}}_p^{(t)}$ are constructed as $\tilde{\mathbf{A}}_{\!BU,p} (\Delta \boldsymbol{\varphi}^{\!(t)}) \!=\! [\tilde{\mathbf{a}}_{\!BU\!,p}(\varphi_1 \!+\! \Delta \varphi_1^{\!(t)}), \!\cdots\!, \tilde{\mathbf{a}}_{\!BU\!,p}(\varphi_{\!N_U} \!+\! \Delta \varphi_{\!N_U}^{\!(t)}\!)]$, $\mathbf{A}_{\!R} (\!\Delta \bm{\bar{\vartheta}}^{\!(t)}\!, \Delta \mathbf{\bar{r}}^{\!(t)}) \!\!=\! [\mathbf{a}_{R}(\bar{\vartheta}_1 \!+\! \Delta \bar{\vartheta}_1^{\!(t)}, \bar{r}_1 \!+\! \Delta \bar{r}_1^{\!(t)}), \!\cdots\!, \mathbf{a}_{R}(\bar{\vartheta}_{\bar{N}_R} \!+\! \Delta \bar{\vartheta}_{\bar{N}_R}^{\!(t)},  \bar{r}_{\bar{N}_R} \!+\! \Delta \bar{r}_{\bar{N}_R}^{\!(t)})]$ and $\mathbf{A}_f (\Delta \boldsymbol{\tau}^{\!(t)}) \!=\! [\mathbf{a}_{f}(\tau_1 + \Delta \tau_1^{\!(t)}), \!\cdots\!, \mathbf{a}_{f}(\tau_{N_f} \!+\! \Delta \tau_{N_f}^{\!(t)})]$, respectively. Consequently, the sparse representation of $\bm{\mathcal{R}}_{p}^{(t)}$ in the polar-angle-delay domain is expressed as
{\setlength\abovedisplayskip{5pt}
	\setlength\belowdisplayskip{5pt}
\begin{equation}
	\label{eq:sparse repre of cas_chan}
	\!\bm{\mathcal{R}}_{p}^{\!(t)\!} \!\!=\! \!\bm{\mathcal{Z}}^{\!(t)} \!\!\times_{\!1}\! \tilde{\mathbf{A}}_{\!B\!U\!,p} (\!\Delta \!\boldsymbol{\varphi}^{\!(t)}\!) \!\!\times_{\!2}\! \mathbf{A}_{\!R} (\!\Delta \bm{\bar{\vartheta}}^{\!(t)}\!\!\!, \Delta \mathbf{\bar{r}}^{\!(t)}\!) \!\!\times_{\!3}\! \mathbf{A}_{\!f} (\!\Delta \!\boldsymbol{\tau}^{\!(t)}\!), 
\end{equation}}%
where $\bm{\mathcal{Z}}^{(t)} \in \mathbb{C}^{N_U \times \bar{N}_R \times N_f}$ is the sparse cascaded channel in the polar-angle-delay domain with $L^{(t)}$ nonzero entries corresponding to $\{\gamma_{l}^{(t)} \}_{l\in \mathcal{L}^{(t)}}$. By substituting \eqref{eq:sparse repre of cas_chan} into \eqref{eq:four-order received signal}, $\bm{\mathcal{Y}}_p^{(t)}$  can be further expressed as
{\setlength\abovedisplayskip{4pt}
 \setlength\belowdisplayskip{4pt}
\begin{align}
	\bm{\mathcal{Y}}_p^{\!(t)} \!\!=&  \bm{\mathcal{Z}}^{(t)} \!\!\times_{\!1}\!\! \big( \mathbf{X}_p^{\!(t)}  \tilde{\mathbf{A}}_{\!B\!U\!,p} (\!\Delta \bm{\varphi}^{\!(t)}\!) \big) \!\times_{\!2}\!  \big( (\!\mathbf{v}_p^{\!(t)}\!)^{\!T} \mathbf{A}_{\!R} (\Delta \bm{\bar{\vartheta}}^{(t)}\!\!, \Delta \mathbf{\bar{r}}^{(t)}) \big) \nonumber \\
	& \times_3 \big(  \sqrt{p_T}\text{diag}(\mathbf{x}_p)\mathbf{S} \mathbf{A}_f (\Delta \bm{\tau}^{(t)}) \big) + \bar{\bm{\mathcal{N}}}_{p}^{(t)}. 
\end{align}}%
Based on \eqref{eq:vec of tensor}, the  vectorized form of $\bm{\mathcal{Y}}_p^{(t)}$, denoted by $\mathbf{y}_p^{(t)}$, can be expressed as
{\setlength\abovedisplayskip{4pt}
\setlength\belowdisplayskip{4pt}
\begin{align}
	\mathbf{y}_p^{\!(t)} \!\!=\! & \Big[\! \big(  \!\sqrt{p_T}\text{diag}(\mathbf{x}_p)\mathbf{S} \mathbf{A}_{\!f} (\!\Delta \bm{\tau}^{\!(t)}\!)  \big) \!\otimes\! \big((\!\mathbf{v}_p^{(t)}\!)^{\!T} \mathbf{A}_{\!R} (\Delta \bm{\bar{\vartheta}}^{\!(t)}\!\!, \Delta \mathbf{\bar{r}}^{\!(t)}) \big)\nonumber \\
	& \! \otimes\! \big(\mathbf{X}_p^{\!(t)}  \tilde{\mathbf{A}}_{\!BU\!,p} (\Delta \bm{\varphi}^{\!(t)}) \big)   \Big] \mathbf{z}^{\!(t)} \!\!+\! \mathbf{\bar{n}}_p^{(t)} \!\!=\! \mathbf{F}_p^{(t)} \mathbf{z}^{\!(t)} \!+\! \mathbf{\bar{n}}_p^{(t)},
\end{align}}%
where $\mathbf{z}^{\!(t)}  \!\in\!  \mathbb{C}^{\bar{N}  \!\times\!   1}$ and $\mathbf{\bar{n}}_p^{\!(t)}  \!\in\!  \mathbb{C}^{R_{\!B} \bar{K} \!\times\! 1}$ are the  vectorized  forms of  $\bm{\mathcal{Z}}^{\!(t)}$ and $\bar{\bm{\mathcal{N}}}_{p}^{\!(t)}$, respectively, with $\bar{N}\!\! =\!\! N_{\!U}  \bar{N}_{\!R} N_{\!f}$ denoting the dimension of $\mathbf{z}^{(t)}$; $ \mathbf{F}_p^{\!(t)} \!\!\!\triangleq\!\!\! \big(\!   \sqrt{p_T}  \text{diag}(\mathbf{x}_{\!p})  \mathbf{S}  \mathbf{A}_{\!f} (\!\Delta \bm{\tau}^{\!(t)})  \big) \!\otimes \big((\mathbf{v}_p^{(t)})^T  \mathbf{A}_R (\Delta \bm{\bar{\vartheta}}^{(t)}, \Delta \mathbf{\bar{r}}^{(t)}) \big) \otimes \big(\mathbf{X}_p^{(t)}  \tilde{\mathbf{A}}_{BU,p} (\Delta \bm{\varphi}^{(t)}) \big) $ denotes the measurement matrix. By stacking the $P$ received uplink signals at the BS, we can obtain 
\begin{equation}
	\label{observation model}
	\mathbf{y}^{(t)} = \mathbf{F}^{(t)} \mathbf{z}^{(t)} + \mathbf{\bar{n}}^{(t)},
\end{equation}
where  $\mathbf{y}^{\!(t)} \!=\! [\mathbf{y}_1^{\!(t)}; \!\cdots\!; \mathbf{y}_P^{\!(t)}]$, $\mathbf{F}^{(t)} = [\mathbf{F}_1^{(t)}; \!\cdots\! ;  \mathbf{F}_P^{(t)}]$ and $\mathbf{\bar{n}}^{(t)} \!=\! [\mathbf{\bar{n}}_1^{(t)}; \!\cdots\!; \mathbf{\bar{n}}_P^{(t)}]$.

\subsection{Spatially Structured Sparse Probability Model for CE}
\label{subsec: spa-struct-sps model for CE}
\par To characterize the spatially structured sparsity inherent in the sparse cascaded channel $\mathbf{z}^{(t)}$ during phase I (i.e., $t=1$), we propose a hierarchical prior model for sparse Bayesian CE. For conciseness, we omit the superscript $(t)$ and focus on the initial frame in the CE phase. In particular, let $\bm{\nu} \triangleq [\nu_1, \cdots, \nu_{\bar{N}}]^T $,  $\bm{\omega} \triangleq [\omega_1, \cdots, \omega_{\bar{N}}]^T$ and
$\mathbf{s} \triangleq [s_1, \cdots, s_{\bar{N}}]^T $ denote the magnitude, phase and support vector of $\mathbf{z}$, respectively, where $s_n = 1$, $\forall n \in \bar{\mathcal{N}} \triangleq \{1, \cdots, \bar{N}\}$, indicates the presence of an active path at the $u$-th angle grid point, the $m$-th position grid point and the $k$-th delay grid point, with the mapping from the three-dimensional (3D) index $(u, m, k)$ to the linear index $n$ defined by $n = (k-1)\bar{N}_R N_U + (m-1)N_U + u$. Furthermore, define $\mathbf{s}_U \triangleq [s_{U,1}, \cdots, s_{U,N_U}]^T$, $\mathbf{s}_R \triangleq [s_{R,1}, \cdots, s_{R,\bar{N}_R}]^T$ and $\mathbf{s}_K \triangleq [s_{K,1}, \cdots, s_{K,N_f}]^T$ as the angle, position, and delay support vectors for the three modes of $\bm{\mathcal{Z}}$, respectively, where $s_{U,u} = 1$, if $\exists s_{u,m,k} = 1$,\footnote{For convenience, we replace the linear index $n$ with the 3D index $(u,m,k)$ for sparse Bayesian CE during phase I, while retaining the linear index $n$ for sparse Bayesian CT during phase II.} $ \forall m,k$, and $s_{U,u} = -1$ otherwise, $\forall u \in \mathcal{N}_U$; $s_{R,m} = 1$, if $\exists s_{u,m,k} = 1$, $\forall u,k$, and $s_{R,m} = -1$ otherwise, $\forall m \in \bar{\mathcal{N}}_R$;  $s_{K,k} = 1$, if $\exists s_{u,m,k} = 1$, $\forall u,m$, and $s_{K,k} = -1$ otherwise, $\forall k \in \mathcal{N}_f$. Then, the spatially structured sparse prior distribution, which incorporates the DFO $f_d$ and the set of off-grid vectors (denoted by $\bm{\Xi} \triangleq \{ \Delta \bm{\varphi}, \Delta\bm{\tau}, \Delta\bar{\bm{\vartheta}}, \Delta\bar{\mathbf{r}} \}$), is given by
\begin{align}
	\label{spa-struct-sps prior}
	p(\bm{\nu}, \bm{\omega},& \mathbf{s}, \mathbf{s}_U, \mathbf{s}_R, \mathbf{s}_K, f_d, \bm{\Xi}) = p(\bm{\nu}|\mathbf{s}) p(\bm{\omega}|\mathbf{s}) \nonumber \\
	& p(\mathbf{s}|\mathbf{s}_U, \mathbf{s}_R, \mathbf{s}_K) p(\mathbf{s}_U)  p(\mathbf{s}_R) p(\mathbf{s}_K)  p(f_d) p(\bm{\Xi}).
\end{align}
\subsubsection{Probability Model for Channel Amplitude $\bm{\nu}$ and Phase $\bm{\omega}$}
Notice that the $l$-th cascaded path gain $\gamma_l = \alpha_l  \beta$, corresponding to a non-zero element of $\mathbf{z}$, is derived from  the product of two independent complex path gains $\alpha_l$ and $\beta$, each modeled as a circularly symmetric complex Gaussian (CSCG) variable \cite{zhou2022irsce,wang2021joint}, i.e., $\alpha_l \sim \mathcal{CN}(0, \sigma_{H,l}^2)$ and $\beta \sim \mathcal{CN}(0, \sigma_{G}^2)$, with $\sigma_{H,l}^2$ and $\sigma_{G}^2$ being the associated large-scale path losses. Consequently, $\gamma_l$ follows a complex double Gaussian distribution \cite{Donoughue2012CNN}, represented as $\gamma_l \sim \mathcal{CNN}(0,\sigma_{H,l}^2; 0, \sigma_{G}^2)$, and the marginal distributions for its magnitude and phase, denoted by $f_{A}(|\gamma_l|)$  and $f_{P}(\angle \gamma_l)$, are defined as follows \cite{Donoughue2012CNN}:
\begin{align}
	\label{marg-distri CNN}
	f_{A}(|\gamma_l|) \!=\! \frac{4|\gamma_l|}{\sigma_{\!H,l}^2 \sigma_{\!G}^2} K_0(\frac{2|\gamma_l|}{\sigma_{\!H,l} \sigma_{\!G}}), \quad f_{P}(\angle \gamma_l) \!\!=\! \mathcal{U}(0,2\pi),
\end{align}
where $K_0$ is the modified Bessel function of the second kind with order zero and  $\mathcal{U}(a,b)$ denotes a uniform distribution from $a$ to $b$. Based on the above definitions, the probability models for the magnitude $\bm{\nu}$ and phase $\bm{\omega}$ of $\mathbf{z}$ are given by

\begin{align}
	\label{prab-model nu-omg}
	\!\!p(\bm{\nu}| \mathbf{s}) \!\!=\!\!\! \prod_{u\!,m\!,k}\!\! p(\nu_{u\!,m\!,k}\!|s_{u\!,m\!,k}), \;\; p(\bm{\omega}| \mathbf{s}) \!\!=\!\!\!\! \prod_{u\!,m\!,k}\!\!\! p(\omega_{u\!,m\!,k}\!|s_{u\!,m\!,k}), \!
\end{align}
where $p(\nu_{u\!,m\!,k}|s_{u\!,m\!,k}) = (1- \mathbb{I}(s_{u\!,m\!,k}))  \delta(\nu_{u\!,m\!,k})   + \mathbb{I}(s_{u,m,k}) \\ f_A(\nu_{u\!,m\!,k})  $ and  $p(\omega_{u\!,m\!,k}|s_{u\!,m\!,k}) \!=\! (1-\mathbb{I}(s_{u\!,m\!,k}))\delta( \omega_{u\!,m\!,k} )  \!+\! \mathbb{I}(s_{u\!,m\!,k}) f_P(\omega_{u\!,m\!,k})   $.

\begin{figure}[t]
    \centering
    \begin{subfigure}{0.7\columnwidth}
        \centering
        \includegraphics[width=\linewidth]{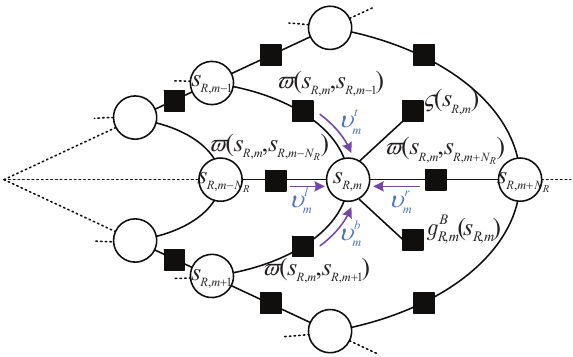}
        \caption{}
        \label{pic:MRF_polar_MF}
    \end{subfigure}
    \hfill
    \begin{subfigure}{0.28\columnwidth}
        \centering
        \includegraphics[width=\linewidth]{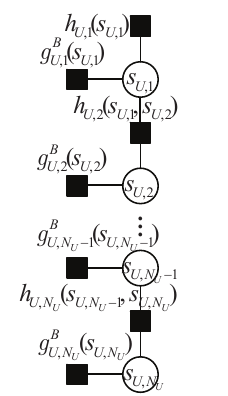}
        \caption{}
        \label{pic:MRF_angle_MC}
    \end{subfigure}
    \caption{Factor graphs of the 4-connected MRF (a) and the Markov chain (b). For clarity, only the factor nodes associated with the central variable node $s_{R,m}$ are plotted in the figure.}
    \label{pic:MC}
    \vspace{-0.2cm}
\end{figure}

\subsubsection{Probability Model for Support Vector $\mathbf{s}$}
The conditional prior distribution $p(\mathbf{s}|\mathbf{s}_{U}, \mathbf{s}_R, \mathbf{s}_K)$ is given by
\begin{align}
	\label{prab-model s,sU,sR,sK}
	p(\mathbf{s}|\mathbf{s}_{U}, \mathbf{s}_R, \mathbf{s}_K) = \prod_{u,m,k} 	p(s_{u,m,k}|s_{U,u}, s_{R,m}, s_{K,k}),
\end{align}
where $p(s_{u,m,k}|s_{U,u}, s_{R,m}, s_{K,k}) = (1 - \mathbb{I}(s_{U,u}, s_{R,m}, s_{K,k}) ) \\ \delta(s_{u,m,k}) + \mathbb{I}(s_{U,u}, s_{R,m}, s_{K,k}) (p_s)^{\mathbb{I}(s_{u\!,m\!,k})} (1-p_s)^{1-\mathbb{I}(s_{u\!,m\!,k})}$ with $p_s$ being the probability of $s_{u\!,m\!,k} = 1$ conditioned on $s_{U,u} = s_{R,m} = s_{K,k} = 1$.

\subsubsection{Probability Model for Support Vector $\mathbf{s}_R$}
To effectively capture the 2D polar-domain block sparsity inherent in the second mode of $\bm{\mathcal{Z}}$, we model the position support vector $\mathbf{s}_R$ using a Markov random field (MRF). Specifically, the Ising model \cite{som2011approximate} is adopted to capture local dependencies among neighboring positions, which provides a soft prior that encourages clustered nonzero elements, i.e., 
\begin{equation}
	\label{Ising model}
	p(\mathbf{s}_R) \!=\! \frac{1}{C} \prod_{m=1}^{\bar{N}_R} \Big[ \varsigma(s_{R,m}) \!\prod_{m^\prime \in \mathcal{D}_m} \! \varpi^{\frac{1}{2}} (s_{R,m},s_{R,m^\prime}) \Big],
\end{equation}
where $C$ is a normalization constant; $\mathcal{D}_m \subset \bar{\mathcal{N}}_R$ represents the set of all neighboring indices of $m$; $\varsigma(s_{R,m}) = e^{-\bar{\eta}_m s_{R,m}}$ and $\varpi  (s_{R,m}, s_{R,m^\prime}) = e^{\check{\eta}_{m,m^\prime} s_{R,m} s_{R,m^\prime} }$ with $\bar{\eta}_m$ and $\check{\eta}_{m,m^\prime}$ being the bias and interaction parameters of the MRF. To intuitively illustrate the internal structure of the MRF, a factor graph of the 4-connected MRF\footnote{In other CE tasks, the sparse channel may exhibit 3D block sparsity \cite{kuai2019struct}, which can be characterized using a 8-connected MRF.} is depicted in Fig. \ref{pic:MC}(\subref{pic:MRF_polar_MF}), which is able to characterize the 2D polar-domain block sparsity. By adjusting $\bar{\eta}_m$ and $\check{\eta}_{m,m^\prime}$, we can effectively control the degree of sparsity and the block size of $\mathbf{s}_R$. Specifically, increasing $\bar{\eta}_m$ enhances the sparsity of $\mathbf{s}_R$, while increasing $\check{\eta}_{m,m^\prime}$ results in larger non-zero blocks. It is noteworthy that in scenarios involving XL-IRS, adjusting the substantial number of $\bar{\eta}_m$ and $\check{\eta}_{m,m^\prime}$ is intractable. To address this issue, we propose a simplified method to design $\bar{\eta}_m$ and $\check{\eta}_{m,m^\prime}$, which is based on the constructed position grid $[\bar{\bm{\vartheta}}, \bar{\mathbf{r}}]$ in the polar domain. In particular, $\bar{\eta}_m$ and $\check{\eta}_{m,m^\prime}$ are given by
$\bar{\eta}_m = \bar{\eta} (1-\bar{\vartheta}_m^2)/q_m$ and
$\check{\eta}_{m,m^\prime} = \check{\eta}(\bar{\vartheta}_m^2 q_m + \bar{\vartheta}_{m^\prime}^2 q_{m^\prime})/2$, respectively, where $\bar{\eta}$ and $\check{\eta}$ are the global bias and interaction parameters, respectively, and $q_m = \mod(m-1,S)+1$ denotes the index of the distance grid point for the $m$-th position. Adopting the proposed method for $\bar{\eta}_m$ and $\check{\eta}_{m,m^\prime}$ results in weaker sparsity and larger non-zero blocks near the pole, while stronger sparsity with smaller non-zero blocks away from the pole, thus accurately capturing the expected 2D polar-domain block sparsity with only two adjustable parameters (i.e., $\bar{\eta}$ and $\check{\eta}$).

\subsubsection{Probability Model for Support Vectors $\mathbf{s}_U$ and $\mathbf{s}_K$}
Since the limited and clustered scatterers lead to 1D clustered sparsity in both the angle and delay domains of $\bm{\mathcal{Z}}$, we propose to utilize a Markov chain, as depicted in Fig. \ref{pic:MC}(\subref{pic:MRF_angle_MC}),  to characterize the angle support vector $\mathbf{s}_U$ and the delay support vector $\mathbf{s}_K$, i.e., 
\begin{align}
	&p(\mathbf{s}_U) = p(s_{U,1})\prod_{u=2}^{N_U}p(s_{U,u}|s_{U,u-1}), \label{markov chain sU} \\
	&p(\mathbf{s}_K) = p(s_{K,1})\prod_{k=2}^{N_f}p(s_{K,k}|s_{K,k-1}), \label{markov chain sK}
\end{align}
with the transition probability  $p(s_{U,u} \!= \! 1|s_{U,u-1} \!= \! -1) = \lambda_{-1,1}^U$, $p(s_{U,u} \!= \! -1|s_{U,u-1}\!=\!1)\! =\! \lambda_{1,-1}^U$, $p(s_{K,k}\!=\!1|s_{K,k-1}\!=\!-1) \!=\! \lambda_{-1,1}^K$ and $p(s_{K,k}\!=\!-1|s_{K,k-1}\!=\!1) \!=\! \lambda_{1,-1}^K$. The initial distributions $p(s_{U,1})$ and $p(s_{K,1})$ are set to the steady-state distributions of the Markov chains in \eqref{markov chain sU} and \eqref{markov chain sK}, respectively, i.e., $p(s_{U,1}) = \frac{\lambda_{-1,1}^U}{\lambda_{-1,1}^U + \lambda_{1,-1}^U}$ and $p(s_{K,1}) = \frac{\lambda_{-1,1}^K}{\lambda_{-1,1}^K + \lambda_{1,-1}^K}$.

\subsubsection{Probability Model for DFO $f_d$ and Off-Grid Vectors $\bm{\Xi}$}
Assuming that the maximum user speed $v_{max}$ is known as a prior information\footnote{If the true maximum user speed is unknown, $f_{d,\max}$ can be chosen as a sufficiently large value based on the coarse estimate obtained in stage I, which will be discussed in Section ~\ref{sec:TS-CET_Algorithm}.}, for the DFO $f_d$, i.e. $p(f_d) \!=\! \mathcal{U}(0,f_{d,max})$, with $f_{d,max} \!=\! \frac{v_{max}}{\lambda_c}$, and $p(\bm{\Xi})$ is modeled by an independent and identically distributed (i.i.d.) Gaussian prior \cite{xu2023successive} as
\begin{align}
	\label{prab-model Xi}
	p(\bm{\Xi}) = \prod_{u,m,k} p(\Delta \varphi_u) p(\Delta \bar{\vartheta}_m) p(\Delta \bar{r}_m) p(\Delta \tau_k),
\end{align}
where $p(\Delta \varphi_u) = \mathcal{N}(0, \varrho_{\varphi}^2)$, $p(\Delta \bar{\vartheta}_m) = \mathcal{N}(0, \varrho_{\bar{\vartheta}}^2)$, $p(\Delta \bar{r}_m) = \mathcal{N}(0, \varrho_{\bar{r}}^2)$  and  $p(\Delta \tau_k) =  \mathcal{N}(0, \varrho_{\tau}^2)$ with $\varrho_{\varphi}^2$, $\varrho_{\bar{\vartheta}}^2$, $\varrho_{\bar{r}}^2$ and $\varrho_{\tau}^2$ being the corresponding off-grid variances.

\subsection{Temporally Dynamic Sparse Probability Model for CT}
\label{Temporally Dynamic Sparse Probability Model}
To characterize the temporally dynamic sparsity inherent in the sparse cascaded channel $\mathbf{z}^{(t)}$ during phase II (i.e., $t \geq 2$), we propose a Markov prior model for sparse Bayesian CT. For ease of notation, let $\bm{\nu}^{(2:T)}$ denote a time series of the sparse channel  amplitudes $\{\bm{\nu}^{(2)}, \!\cdots\!, \bm{\nu}^{(T)} \}$ ($\bm{\omega}^{(2:T)}$, $\mathbf{s}^{(2:T)}$, $f_d^{(2:T)}$, $\bm{\Xi}^{(2:T)}$, $\Delta \bm{\varphi}^{(2:T)}$, $\Delta \bar{\bm{\vartheta}}^{(2:T)}$, $\Delta \bar{\mathbf{r}}^{(2:T)}$ and $\Delta {\bm{\tau}}^{(2:T)}$ are similarly defined). Then, the temporally dynamic sparse prior distribution is given by
\begin{align}
	\label{tem-dynm-sps prior}
	&\!\!p(\bm{\nu}^{\!(\!2:T\!)}\!\!,\bm{\omega}^{\!(\!2:T\!)}\!\!,\mathbf{s}^{\!(\!2:T\!)}\!\!,f_d^{\!(\!2:T\!)}\!\!,\bm{\Xi}^{\!(\!2:T\!)})\!\!=\! p(\bm{\nu}^{\!(\!2:T\!)}|\mathbf{s}^{\!(\!2:T\!)})p(\bm{\omega}^{\!(\!2:T\!)}|\mathbf{s}^{\!(\!2:T\!)}) \nonumber \\ 
	&\!\!p(\!\mathbf{s}^{(\!2:T\!)}\!)p(\!f_d^{(\!2:T\!)}\!)p(\!\Delta \bm{\varphi}^{\!(\!2:T\!)}\!)p(\!\Delta \bar{\bm{\vartheta}}^{(\!2:T\!)}\!)p(\Delta \bar{\mathbf{r}}^{(\!2:T\!)}\!)p(\Delta {\bm{\tau}}^{(\!2:T\!)}\!).
\end{align}

\subsubsection{Probability Model for Channel Amplitude $\bm{\nu}^{(2:T)}$ and Phase $\bm{\omega}^{(2:T)}$}
Since the user moves at a low speed, both the channel magnitude and phase evolve smoothly over time, exhibiting a temporal structure that can be exploited. As such, we can exploit the Gauss-Markov process  \cite{ziniel2012dynamic} to capture the temporal correlation of $\nu_n^{{\!\!(t)}}$ and $\omega_n^{{\!\!(t)}}$ ($\forall t \!\in\!\! \{2, \!\cdots\!,\!T\}$\!, $\!\forall  n \!\!\in\! \bar{\mathcal{N}}$) as
\begin{align}
	\label{Gaus-Makv nu-omg}
	&\nu_{n}^{(t)} = (1-\chi_{\nu})(\nu_{n}^{(t-1)} - \mu_{\nu} ) + \chi_{\nu}\varepsilon_{\nu,n}^{(t)} + \mu_{\nu}, \\
	&\omega_{n}^{(t)} = (1-\chi_{\omega})(\omega_{n}^{(t-1)} - \mu_{\omega} ) + \chi_{\omega}\varepsilon_{\omega,n}^{(t)} + \mu_{\omega},
\end{align}
where $\chi_{\nu}$ and $\chi_{\omega}$, within the range $[0,1]$, control the temporal correlation; 
$\mu_{\!\nu}$ and $ \mu_{\!\omega}$ represent the steady-state means of the process; $\varepsilon_{\nu,n}^{(t)} \sim \mathcal{N}(0, \zeta_{\nu})$ and $\varepsilon_{\omega,n}^{(t)}\sim \mathcal{N}(0, \zeta_{\omega})$ are Gaussian perturbations with variances $\zeta_{\nu}$ and $\zeta_{\omega}$, respectively. Consequently, the conditional prior distributions $p(\bm{\nu}^{(2:T)}|\mathbf{s}^{(2:T)})$ and $p(\bm{\omega}^{(2:T)}|\mathbf{s}^{(2:T)})$ can be expressed as
\begin{align}
	\label{prab-model nu omg}
	& p(\bm{\nu}^{(2:T)}|\mathbf{s}^{(2:T)}) \!\!=\!\! \prod_{n=1}^{\bar{N}}\!p(\nu_n^{(2)}|s_n^{(2)})\prod_{t=3}^{T}\!p(\nu_n^{\!(t)}|\nu_n^{\!(t\!-\!1)}\!, s_n^{\!(t)}), \\
	&p(\bm{\omega}^{(2:T)}|\mathbf{s}^{(2:T)}) \!\!=\!\! \prod_{n=1}^{\bar{N}}\!p(\omega_n^{(2)}|s_n^{(2)})\prod_{t=3}^{T}\!p(\omega_n^{\!(t)}|\omega_n^{\!(t\!-\!1)}\!, s_n^{\!(t)}),
\end{align}
where $p(\nu_n^{\!(t)}|\nu_n^{\!(t\!-\!1\!)}\!\!, s_n^{\!(t)}) \!\!=\!\! (1\!-\!\mathbb{I}(s_n^{\!(t)})) \delta(\nu_n^{\!(t)}) \!+\! \mathbb{I}(s_n^{\!(t)})p(\nu_n^{\!(t)}|\nu_n^{\!(t\!-\!1)})$, $p(\omega_n^{\!(t)}|\omega_n^{\!(t-1)}\!, s_n^{\!(t)}) \!=\! (1-\mathbb{I}(s_n^{\!(t)})) \delta(\omega_n^{\!(t)}) + \mathbb{I}(s_n^{\!(t)})p(\omega_n^{\!(t)}|\omega_n^{(t-1)})$, $p(\nu_n^{\!(t)}|\nu_n^{\!(t\!-\!1\!)}\!) \!\!=\!\! \mathcal{N}( (1\!-\!\chi_{\nu})\nu_n^{\!(t\!-\!1\!)} \!+\!  \chi_{\nu}\mu_{\nu},  \chi_{\nu}^2 \zeta_{\nu} )$ and $p(\omega_n^{\!(t)}|\omega_n^{\!(t\!-\!1\!)}\!) \!= \mathcal{N}( (1-\chi_{\omega})\omega_n^{\!(t\!-\!1)} +  \chi_{\omega}\mu_{\omega},  \chi_{\omega}^2 \zeta_{\omega} )$. The initial distributions $p(\nu_n^{\!(2)}|s_n^{\!(2)})$ and $p(\omega_n^{\!(2)}|s_n^{\!(2)})$ in the CT phase can be derived from the CE results $\{\!\hat{\nu}_n, \hat{\omega}_n\!\}$ as $p(\nu_n^{\!(2)}|s_n^{\!(2)}\!) \!\!=\!\! (\!1\!-\! \mathbb{I}(s_n^{\!(2)}\!))\delta(\nu_n^{\!(2)}\!) + \mathbb{I}(s_n^{\!(2)}\!) \hat{p}(\nu_n^{\!(2)}\!) $ with $ \hat{p}(\nu_n^{\!(2)}) \!=\!  \mathcal{N}( (1\!-\!\chi_{\nu}) \hat{\nu}_n \!+\!  \chi_{\nu}\mu_{\nu},  \chi_{\nu}^2 \zeta_{\nu} )$, and $p(\omega_n^{\!(2)}|s_n^{\!(2)}) = (1- \mathbb{I}(s_n^{\!(2)}))\delta(\omega_n^{\!(2)}) + \mathbb{I}(s_n^{\!(2)}) \hat{p}(\omega_n^{\!(2)}) $ with $\hat{p}(\omega_n^{\!(2)}) \!= \mathcal{N}( (1-\chi_{\omega})\hat{\omega}_n +  \chi_{\omega}\mu_{\omega},  \chi_{\omega}^2 \zeta_{\omega} )$, respectively.

\subsubsection{Probability Model for Support Vector $\mathbf{s}^{(2:T)}$}
Due to the slowly varying propagation environment and the user's low-speed movement, the support vector $\mathbf{s}^{\!(t)}$ also evolves slowly over time. Thus, we model the temporal correlation of $s_n^{\!(t)}$  ($\forall t \!\in\!\! \{2, \!\cdots\!, T\}$, $\!\forall n \!\in\! \bar{\mathcal{N}}$) using a Markov chain as 
\begin{align}
	\label{prab-model s}
	p(\mathbf{s}^{(2:T)}) = \prod_{n = 1}^{\bar{N}}p(s_n^{(2)})\prod_{t=3}^Tp(s_{n}^{\!(t)}|s_{n}^{(t-1)}),
\end{align}
with the transition probability $p(s_{n}^{\!(t)} \!\!=\! 1|s_{n}^{\!(t\!-\!1\!)} \!\!=\! -1) \!=\! \lambda_{-\!1,1}$ and $p(s_{n}^{\!(t)} \!\!=\! -\!1|s_{n}^{\!(t\!-\!1\!)} \!\!=\!\! 1) \!\!=\!\! \lambda_{1,-\!1}$. The initial distribution $p(s_n^{(2)})$ is given by $p(\!s_n^{\!(2)}) \!\!=\!\! (\!1\!-\!\mathbb{I}({\hat{s}_n}\!)) \lambda_{-\!1,1}^{ \mathbb{I}(\!{s_n^{\!(2)}}\!) }  (\!1\!-\!\lambda_{-\!1,1\!})^{1\!-\!\mathbb{I}(\!{s_n^{\!(2)\!}}) \!} \!+\! \mathbb{I}({\hat{s}_n}\!)  \lambda_{1,-\!1}^{\! 1\!-\!\mathbb{I}(\!{s_n^{\!(2)\!}}) }  \\ (1\!-\!\lambda_{1,-\!1})^{\mathbb{I}({s_n^{\!(2)\!}})} $ with $\hat{s}_n$ being the CE result.

\subsubsection{Probability Model for DFO $f_d^{(2:T)}$}
Assuming that the user's acceleration is small, then we can infer that the DFO $f_d^{\!(t)}$ exhibits a high degree of temporal structure that can be leveraged. Similarly, the Gauss-Markov process is applied to capture the temporal correlation of $f_d^{\!(t)}$ ($\forall t \in \{2, \cdots, T \}$) as 
\begin{align}
	\label{prab-model fd}
	f_d^{\!(t)} = (1-\chi_{f})(f_d^{(t-1)} - \mu_{f} ) + \chi_{f}\varepsilon_{f}^{\!(t)} + \mu_{f},
\end{align}
resulting  in  the prior distribution $p(f_d^{(2:T)}) \,=\, p(f_d^{(2)})\,\prod_{t=3}^T \\ p(f_d^{(t)} | f_d^{(t-1)})$ with $p(f_d^{(t)} | f_d^{(t-1)}) =  \mathcal{N}( (1-\chi_{f})f_d^{(t-1)} \!+\!  \chi_{f}\mu_{f},  \chi_{f}^2 \zeta_{f} )$, where $\chi_{f}$, $\mu_{f}$, $\varepsilon_{f}^{\!(t)}$ and $\zeta_{f}$ are defined similarly to $\chi_{\nu}$, $\mu_{\nu}$, $\varepsilon_{\nu,n}^{\!(t)}$ and $\zeta_{\nu}$ mentioned above. The initial distribution $p(f_d^{(2)})$ is given by $p(f_d^{(2)}) =  \mathcal{N}( (1\!-\!\chi_{f}) \hat{f}_d \!+\!  \chi_{f}\mu_{f},  \chi_{f}^2 \zeta_{f} )$ with $\hat{f}_d$ being the CE result.

\subsubsection{Probability Model for Off-Grid Vectors $\bm{\Xi}^{(2:T)}$}
Similar to the other channel parameters, the off-grid vectors evolve smoothly over time and can be formulated based on their respective temporal correlations as \cite{Zhang2016tracking}
\begin{align}
	\label{Rand-Gaus offgrid}
	&\Delta \bm{\varphi}^{\!(t)} = \Delta \bm{\varphi}^{(t-1)} + \mathbf{c}_{\bm{\varphi}}^{\!(t)}, \quad 
	\Delta \bm{\tau}^{\!(t)} = \Delta \bm{\tau}^{(t-1)} + \mathbf{c}_{\bm{\tau}}^{\!(t)}, \\
	& \Delta \bar{\bm{\vartheta}}^{\!(t)} = \Delta \bar{\bm{\vartheta}}^{(t-1)} + \mathbf{c}_{\bar{\bm{\vartheta}}}^{\!(t)}, \quad \Delta \bar{\mathbf{r}}^{\!(t)} = \Delta \bar{\mathbf{r}}^{(t-1)} + \mathbf{c}_{\bar{\mathbf{r}}}^{\!(t)},
\end{align}
where $\mathbf{c}_{\bm{\varphi}}^{\!(t)} \sim \mathcal{N}(\mathbf{0}, \zeta_{\varphi}\mathbf{I})$, $\mathbf{c}_{\bm{\tau}}^{\!(t)} \sim \mathcal{N}(\mathbf{0}, \zeta_{\tau}\mathbf{I})$, $\mathbf{c}_{\bar{\bm{\vartheta}}}^{\!(t)} \sim \mathcal{N}(\mathbf{0}, \zeta_{\bar{\vartheta}}\mathbf{I})$ and $\mathbf{c}_{\bar{\mathbf{r}}}^{\!(t)} \sim \mathcal{N}(\mathbf{0},\zeta_{\bar{r}}\mathbf{I})$. Consequently, the prior distribution for $\Delta \bm{\varphi}^{(2:T)}$ can be expressed as
\begin{align}
	\label{prab-model Delta varphi}
	 p(\Delta \bm{\varphi}^{(2:T)}) = \prod_{u=1}^{N_U} p(\Delta \varphi_u^{(2)}) \prod_{t=3}^T p(\Delta \varphi_u^{\!(t)} | \Delta \varphi_u^{(t-1)}),
\end{align}
where $p(\Delta \varphi_u^{\!(t)} | \Delta \varphi_u^{(t-1)}) = \mathcal{N}(\Delta \varphi_u^{(t-1)}, \zeta_{\varphi})$ and the initial distribution $p(\Delta \varphi_u^{(2)})$ is given by $p(\Delta \varphi_u^{(2)}) = \mathcal{N}(\Delta \hat{\varphi}_u, \zeta_{\varphi})$ with $\Delta \hat{\varphi}_u$ being the CE result. The prior distributions for $\Delta \bm{\tau}^{(2:T)}$, $\Delta \bar{\bm{\vartheta}}^{(2:T)}$ and $\Delta \bar{\mathbf{r}}^{(2:T)}$ are defined similarly.

\subsection{Active and Passive Beamforming Design for CET}
Since the active beamforming vector/matrix $\mathbf{f}_p^{\!(t)}$ and $\mathbf{W}_p^{\!(t)}$, as well as the passive beamforming vector $\mathbf{v}_p^{\!(t)}$, have significant effects on the performance of CET, it is essential to design them appropriately in both phase I and phase II based on the available prior information. First, we notice that the spatial angle $\vartheta_B$ between the XL-IRS and BS is known and remains unchanged during both phases. Therefore, to achieve active beamforming gain at the BS, $\mathbf{W}_p^{\!(t)}$ ($t = 1, 2, \cdots$) is designed to align with $\vartheta_B$, i.e., $[\mathbf{W}_p^{(t)}]_{:,n} = e^{j\theta_{n,p}^W} \mathbf{a}_{B}(\vartheta_B)$ for $n = 1,\ldots,R_B$ and $p = 1,\ldots,P$, where $\theta_{n,p}^W$ is randomly generated from $[0,2\pi]$. Next, we design $\mathbf{v}_p^{(t)}$ and $\mathbf{f}_p^{(t)}$ differently in phases I and II, reflecting the varying availability of prior information; details are provided in Appendix A.

\section{TS-CET Algorithm}
\label{sec:TS-CET_Algorithm}
\par In the $t$-th frame, given the observations $\mathbf{y}^{(1:t)}$, we aim to compute the minimum mean-squared error (MMSE) estimates of channel magnitudes $\{\nu_n^{(t)}\}$ and phases $\{\omega_n^{(t)}\}$, DFO $f_d^{(t)}$ and off-grid values $\{\Xi_{u,m,k}^{(t)} \}\triangleq \{\Delta \varphi_u^{(t)}, \Delta \bar{\vartheta}_m^{(t)}, \Delta \bar{r}_m^{(t)},  \Delta \tau_k^{(t)}\}$, leveraging both the spatially structured sparsity and temporally dynamic sparsity inherent in the sparse cascaded channel $\mathbf{z}^{(t)}$. Specifically, for $\nu_n^{(t)}$ (similarly for the other variables), $\forall n \in \bar{\mathcal{N}}$,  the MMSE estimates, denoted by $\hat{\nu}_n^{(t)}$,  are derived as 
$\hat{\nu}_n^{(t)} = \mathbb{E}\big[ \nu_n^{(t)} | \mathbf{y}^{(1:t)}  \big]$ 
with the expectation calculated over the marginal posterior 
\begin{align}
	\label{exp:marginal posterior}
	p(\nu_n^{(t)} | \mathbf{y}^{(1:t)} ) \propto \int_{\sim \nu_n^{(t)}} p( \mathbf{y}^{(1:t)}, \bm{\kappa}^{(t)}),
\end{align}
where $\bm{\kappa}^{(t)}\!\! \triangleq\!\! \{ \bm{\nu}, \bm{\omega}, \mathbf{s}, \mathbf{s}_{U}\!, \mathbf{s}_{R}, \mathbf{s}_{K}\!, f_d, \bm{\Xi}  \}$ for  $t\!\!=\!\!1$ and is adjusted to $\bm{\kappa}^{(t)} \! \triangleq \! \{ \bm{\nu}^{(t)}\!\!, \bm{\omega}^{(t)}\!, \mathbf{s}^{(t)}\!, f_d^{(t)}\!, \bm{\Xi}^{(t)} \! \}$ for $t \!\geq\! 2$, and $\int_{\!\sim \nu_n^{\!(t)}}$ denotes the integration over all  variables except $\nu_n^{(t)}$. However, calculating the accurate marginal posterior in \eqref{exp:marginal posterior} is intractable due to the multi-dimensional integration and the presence of loops in the corresponding factor graph.
Although several methods have been proposed to approximate marginal posterior of $\mathbf{z}^{(t)}$ based on the observation model \eqref{observation model}, such as VBI \cite{liu2020tvbi} and sparse Bayesian learning (SBL) \cite{dai2018sbl}, directly applying these methods to the considered CET problem encounters significant challenges: 1) the prior distribution of $\mathbf{z}^{(t)}$ is limited to certain conjugate priors (e.g., the Gaussian and Gamma distributions), hindering the use of more accurate priors (e.g., the complex double Gaussian distribution in \eqref{marg-distri CNN}), 2) the EM framework is required to obtain the maximum likelihood (ML) estimates of the parameters in the measurement matrix $\mathbf{F}^{(t)}$ (e.g., $f_d^{(t)}$ and $\Delta \bm{\varphi}$), which, however, fails to provide the marginal posteriors of these parameters, and 3) there is a lack of algorithms that are able to effectively solve both the CE problem with spatially structured sparsity and the CT problem with temporally dynamic sparsity. To address these challenges, we propose in this section the TS-CET algorithm, which combines the tensor-based OMP, particle-based VBI and message passing methods to provide high-performance approximate marginal posteriors for all the variables in $\bm{\kappa}^{(t)}$.

\begin{figure}[tbp]
	\setlength{\abovecaptionskip}{-0.0cm}
	\centering
	{\includegraphics[width=0.52\textwidth]{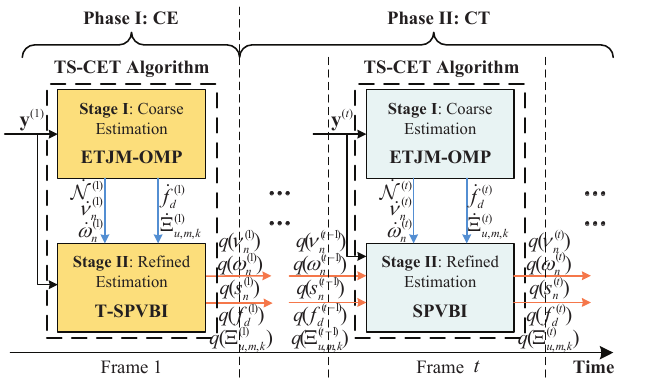}}
	\caption{Flow chart of the proposed TS-CET algorithm.}
	\label{pic:algorithm framework}
	\vspace{-0.3cm}
\end{figure}

\subsection{Outline of the TS-CET Algorithm}
\par Inspired by particle-based methods \cite{zhou2017pvbi} that employ a series of discrete particles with associated weights to sample the intractable posterior distribution, the core of the proposed TS-CET algorithm is the stochastic particle-based VBI (SPVBI) method (as will be elaborated in Section \ref{subsec: T-SPVBI}), which merges the strengths of VBI and particle-based methods. Despite the adaptability of the SPVBI method to a wide range of prior distributions, effectively characterizing the marginal posteriors of all variables in $\bm{\kappa}^{(t)}$ necessitates a large number of particles. However, storing and updating such a substantial number of particles is impractical. To address this issue, the proposed TS-CET algorithm is structured into two stages, as depicted in Fig. \ref{pic:algorithm framework}, with a brief overview of each stage provided below:
\begin{figure*}[b]
	\normalsize
	\vspace*{-8pt}
	\hrulefill
	\vspace*{-4pt}
	\begin{equation}
	\label{joint distb dicomp}
	\begin{aligned}
		p(\mathbf{y}^{\!(\!1:t)}\!\!, \bm{\kappa}^{\!(t)}\!)  \!\!=&  \underbrace{p(\mathbf{y}^{\!(t)} | \bm{\nu}^{\!(t)}\!\!,  \bm{\omega}^{\!(t)}\!\!, f_d^{\!(t)}\!\!, \bm{\Xi}^{\!(t)}  \!) }_{\text{Likelihood function}}
		\int_{\!\bm{\kappa}^{\!(t\!-\!1)}}\! 
		\underbrace{p( \bm{\nu}^{\!(t)}  | \bm{\nu}^{\!(t\!-\!1)}\!\!, \mathbf{s}^{\!(t)}  ) 
		p( \bm{\omega}^{\!(t)}  | \bm{\omega}^{\!(t\!-\!1)}\!\!, \mathbf{s}^{\!(t)}  ) 	
		p(   \mathbf{s}^{\!(t)} |  \mathbf{s}^{\!(t\!-\!1)}   )  
		p(   f_d^{\!(t)} |  f_d^{\!(t\!-\!1)}   ) 
		p(   \bm{\Xi}^{\!(t)}  |  \bm{\Xi}^{\!(t\!-\!1)}   )}_{\text{Temporally dynamic sparse probability model}}  \\
		&\underbrace{p(\bm{\nu}^{\!(t\!-\!1)} \!| \mathbf{y}^{\!(1:t\!-\!1)}\!) 
		p(\bm{\omega}^{\!(t\!-\!1)} \!| \mathbf{y}^{\!(1:t\!-\!1)}\!)
		p(\mathbf{s}^{\!(t\!-\!1)} \!| \mathbf{y}^{\!(1:t\!-\!1)}\!) 
		p(f_d^{(t\!-\!1)} \!| \mathbf{y}^{\!(1:t\!-\!1)}\!)
		p(\bm{\Xi}^{\!(t\!-\!1)} \!| \mathbf{y}^{\!(1:t\!-\!1)}\!)}_{\text{Marginal posteriors of $\bm{\kappa}^{(t-1)}$}}.
	\end{aligned}
	\end{equation}
\end{figure*}
\begin{itemize}
	\item  \textbf{Stage I:} Given the observation $\mathbf{y}^{(t)}$, we employ the efficient tensor-based joint mode OMP (ETJM-OMP) method to obtain an estimated non-zero element set $\dot{\mathcal{N}}^{(t)}$ of $\mathbf{z}^{(t)}$ and coarse point estimates $\{ \dot{\nu}_n^{(t)}, \dot{\omega}_n^{(t)}, \dot{f}_d^{(t)}, \dot{\Xi}_{u,m,k}^{(t)} \}$ with low complexity. 
	\item  \textbf{Stage II:} Given the observation $\mathbf{y}^{(t)}$, coarse estimation results $\{\dot{\mathcal{N}}^{(t)}, \dot{\nu}_n^{(t)}, \dot{\omega}_n^{(t)}, \dot{f}_d^{(t)}, \dot{\Xi}_{u,m,k}^{(t)}\}$ from stage I, and prior information \{$q(\nu_n^{(t-1)})$, $q(\omega_n^{(t-1)})$, $q(s_n^{(t-1)})$, $q(f_d^{(t-1)})$,  $q(\Xi_{u,m,k}^{(t-1)})$\} from the previous frame (available only in the CT phase, i.e., $t \geq 2$), we employ the SPVBI method to derive the approximate marginal posteriors \{$q(\nu_n^{(t)})$, $q(\omega_n^{(t)})$, $q(s_n^{(t)})$, $q(f_d^{(t)})$,  $q(\Xi_{u,m,k}^{(t)})$\}, with $q(\nu_n^{(t)}) \approx p(\nu_n^{(t)}|\mathbf{y}^{(1:t)})$.\footnote{The other variables are defined similarly.} Since we only need to calculate the  approximate marginal posteriors associated with the non-zero elements indicated by $\dot{\mathcal{N}}^{(t)}$ at this stage, the memory cost and computational complexity of the SPVBI method are significantly reduced. Furthermore, in the CE phase, we combine the SPVBI and message passing methods via a turbo framework to capture the spatially structured sparsity. While in the CT phase, an appropriate approximation for the joint distribution in \eqref{exp:marginal posterior} is performed (as detailed in Section \ref{subsec: joint distb approx}) to simplify the temporally dynamic message passing, thus allowing the turbo framework to be removed.
\end{itemize} \vspace{-0.2cm}

\subsection{Joint Distribution Approximation in the CT Phase}
\label{subsec: joint distb approx}

In the $t$-th frame of the CT phase, the marginal posteriors of \{$\nu_n^{(t)}$, $\omega_n^{(t)}$, $f_d^{(t)}$, $\Xi_{u,m,k}^{(t)}$\} are calculated based on all the observations $\mathbf{y}^{(1:t)}$. To simplify the temporally dynamic message passing across different frames, we propose to first decompose the joint distribution in \eqref{exp:marginal posterior} into three components, as shown at the bottom of this page, which correspond to the likelihood function of the observation model \eqref{observation model}, the temporally dynamic sparse probability model discussed in Section \ref{Temporally Dynamic Sparse Probability Model}, and the marginal posteriors of $\bm{\kappa}^{(t-1)}$, respectively. 

Then, by substituting the accurate marginal posteriors in the third component with approximate ones based on the outcomes of the SPVBI method, we can express the approximate joint distribution as follows:
\begin{align}
	\label{approx joint dirstb}
	\hat{p}(\mathbf{y}^{(1:t)}\!, \bm{\kappa}^{(t)}) \approx
	& \; p(\mathbf{y}^{(t)} | \bm{\nu}^{(t)},  \bm{\omega}^{(t)}, f_d^{(t)}, \bm{\Xi}^{(t)}  ) 
	\hat{p}(\bm{\nu}^{(t)} | \mathbf{s}^{(t)} ) \nonumber \\
	& \; \hat{p}( \bm{\omega}^{(t)} | \mathbf{s}^{(t)} )
	\hat{p}( \mathbf{s}^{(t)} )
	\hat{p}( f_d^{(t)} )
	\hat{p}(\bm{\Xi}^{(t)}  ),
\end{align}
\par\noindent where $\hat{p}( \bm{\nu}^{\!(t)} \!| \mathbf{s}^{(t)} \!) \!=\! \!\prod_{n=1}^{\bar{N}} \int_{\bm{\nu}^{(t\!-\!1)}}  p(\nu_n^{(t)}|\nu_{n}^{(t\!-\!1)}\!\!, s_n^{(t)} ) \hat{q}(\nu_n^{(t\!-\!1)}) \!= 
\prod_{n=1}^{\bar{N}} (1\!\!-\!\mathbb{I}(s_n^{\!(t)}\!)) \delta(\nu_n^{\!(t)})  \!+\! \mathbb{I}(s_n^{\!(t)})\hat{p}(\nu_n^{(t)}) $,
$\hat{q}(\nu_n^{(t\!-\!1)}) \!=\! \mathcal{N}(\hat{m}_n^{l,(t\!-\!1)}, \\ \hat{\Sigma}_n^{l,(t-1)})$ denotes the Gaussian approximation of the particle-based posterior $q(\nu_n^{(t-1)})$ obtained from the SPVBI method with
$\hat{m}_n^{l,(t\!-\!1)}$ and $\hat{\Sigma}_n^{l,(t\!-\!1)}$ specified later in Section \ref{subsec: T-SPVBI},
and $\hat{p}(\nu_n^{\!(t)}) \!=\! \int_{\bm{\nu}^{(t\!-\!1)}}\! p(\nu_n^{(t)}|\nu_{n}^{(t\!-\!1)}) \hat{q}(\nu_n^{(t\!-\!1)}) \!=\! \mathcal{N}(m_n^{\!l, (t)}\!\!, \Sigma_n^{l, (t)})$ 
with $m_n^{\!l, (t)} \!\!=\!\! (1\!-\!\chi_{\nu}) \hat{m}_n^{l, (t\!-\!1)} \!+\!  \chi_{\nu}\mu_{\nu} $ and $\Sigma_n^{l,(t)} \!=\!(1\!-\!\chi_{\nu})^2 \hat{\Sigma}_n^{l,(t\!-\!1)} \!+\! \\ \chi_{\nu}^2 \zeta_{\nu} $.
 $\hat{p}(\mathbf{s}^{\!(t)}\!) \!\!=\!\! \prod_{n\!=\!1}^{\bar{N}} \! \sum_{\mathbf{s}^{\!(\!t\!-\!1\!)}}\!  p(s_n^{\!(t)} | s_n^{\!(t\!-\!1)} \!) 
q(\!s_n^{\!(t\!-\!1)}\!) \!\!=\!\!  \prod_{n\!=\!1}^{\bar{N}}   \hat{\pi}_n^{\!(t)}   \mathbb{I}(s_n^{\!(t)}) \\ 
\!+\!   (1\!-\!\hat{\pi}_n^{\!(t)}) (1\!-\!\mathbb{I}(s_n^{\!(t)}))  $ with
$\hat{\pi}_n^{\!(t)} \!=\! w_{n,1}^{\!l,(t\!-\!1)}(1\!-\!\lambda_{1\!,-\!1}) \!+\! \lambda_{\!-\!1,1} w_{n,2}^{\!l,(t\!-\!1)}$, and
$\{ w_{n,1}^{\!l,(t\!-\!1)}, w_{n,2}^{\!l,(t\!-\!1)} \}$ specified later.
$\hat{p}( \bm{\omega}^{(t)} | \mathbf{s}^{(t)} )$ is defined similar to $\hat{p}(\bm{\nu}^{(t)} | \mathbf{s}^{(t)} )$, while $\hat{p}( f_d^{(t)} )$ and $\hat{p}(\bm{\Xi}^{(t)}  )$ are defined similar to $\hat{p}(\nu_n^{\!(t)})$. The details are omitted here due to space limitations. Since the approximate priors for the $t$-th frame, i.e.,  \{$\hat{p}(\bm{\nu}^{(t)} | \mathbf{s}^{(t)} )$, $\hat{p}( \bm{\omega}^{(t)} | \mathbf{s}^{(t)} )$, $\hat{p}( \mathbf{s}^{(t)} )$, $\hat{p}( f_d^{(t)} )$,  $\hat{p}(\bm{\Xi}^{(t)}  )$\},  sufficiently exploit the messages passed from previous frames up to $(t-1)$ and effectively perform the temporally dynamic message passing from the $(t-1)$-th to the $t$-th frame, the turbo framework is not required in the CT phase. 

\begin{figure}[tbp]
	\setlength{\abovecaptionskip}{-0.0cm}
	\centering
	{\includegraphics[width=0.45\textwidth]{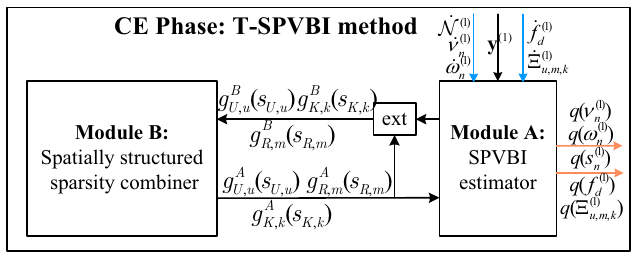}}
	\caption{Top-level diagram of the T-SPVBI method.}
	\label{pic:T_SPVBI}
	\vspace{-0.3cm}
\end{figure}
\subsection{Stage I: ETJM-OMP Method}

\par To provide an estimated non-zero element set $\dot{\mathcal{N}}^{(t)}$ of $\mathbf{z}^{(t)}$ and coarse point estimates $\{ \dot{\nu}_n^{(t)}, \dot{\omega}_n^{(t)}, \dot{f}_d^{(t)}, \dot{\Xi}_{u,m,k}^{(t)}\}$ for stage II, we first introduce the ETJM-OMP method. This method performs a small-scale joint search over the coarse 3D index structure $ \mathcal{S}^{(t)} \triangleq \big( \mathcal{N}_U^{(t)},\, \bar{\mathcal{N}}_R^{(t)},\, \mathcal{N}_f^{(t)} \big)$, where $\mathcal{N}_U^{(t)} \subseteq \mathcal{N}_U$, $\bar{\mathcal{N}}_R^{(t)} \subseteq \bar{\mathcal{N}}_R$, and $\mathcal{N}_U^{(t)} \subseteq \mathcal{N}_f$. In the CE phase, $\mathcal{S}^{(t)}$ is obtained from the tensor-OMP with sequential search (T-OMP-SS) algorithm \cite{Daniel2019tensor}, while in the CT phase, it is determined by the support set $\hat{\mathbf{s}}^{(t-1)}$. Compared with T-OMP-SS, ETJM-OMP improves estimation accuracy by jointly selecting atoms across all tensor dimensions, thereby correcting mismatches caused by sequential search. Moreover, compared with the conventional OMP and tensor-OMP with joint search (T-OMP-JS) algorithms \cite{Daniel2019tensor}, it reduces memory usage and computational complexity. For clarity, we omit the superscript $(t)$ in the sequel of this subsection. To be specific, let $\bm{\mathcal{T}}_p^{(l)}$ denote the residual of  the $p$-th observation ($\forall p \in \mathcal{P}$) at iteration $l$ with $\bm{\mathcal{T}}_p^{(0)} = \bm{\mathcal{Y}}_p$, then the core step of the proposed ETJM-OMP algorithm is summarized as follows 
\begin{equation}
	\label{ETJM-OMP}
	\begin{aligned}
[u^{(l)}, m^{(l)}, k^{(l)}]
= \argmax_{u,m,k} \; \frac{1}{Z} \sum_{p=1}^P 
   | \bm{\mathcal{T}}_p^{(l-1)} 
   \times_{\!1} \mathbf{u}_{U,p}^H(u) 
   \\\times_{\!2} u_{R,p}^H(m)
   \times_{\!3} \mathbf{u}_{K,p}^H(k) |^2,
	\end{aligned}
\end{equation}
where $\mathbf{u}_{K,p}(k) \!=\!\! \sqrt{p_T} \text{diag}(\mathbf{x}_p)\mathbf{S}\mathbf{a}_f(\tau_{k}) $, where $\mathbf{a}_f(\tau_{k})$ corresponds to the $k$-th column of $\mathbf{A}_{\!f}\!(\mathbf{0})$. Similarly, $\mathbf{u}_{U,p}(u) = \mathbf{X}_p \tilde{\mathbf{a}}_{BU,p}(\varphi_{u})$, where $\tilde{\mathbf{a}}_{BU,p}(\varphi_{u})$ corresponds to the $u$-th column of $\tilde{\mathbf{A}}_{\!B\!U\!,p}(\mathbf{0})$, and $u_{R,p}(m) = \mathbf{v}_p^T \mathbf{a}_R(\bar{\vartheta}_{m}, \bar{r}_{m})$ with $ \mathbf{a}_R(\bar{\vartheta}_{m}, \bar{r}_{m})$ denoting the $m$-th column of $\mathbf{A}_{R}(\mathbf{0},\mathbf{0})$. The normalization factor is defined as $Z = \sum_{\!p} \| u_{R,p}(m)  \mathbf{u}_{\!K\!,p}(k^{(l)})   \otimes \mathbf{u}_{U,p}(u^{(l)})  \|^{2} $.  Based on the above joint mode scanning, we can obtain the 3D index $(u^{(l)}, m^{(l)}, k^{(l)})$ of the identified active path corresponding to the three modes of $\bm{\mathcal{Z}}$. By defining the estimated non-sparse measurement matrix as $\mathbf{U}^{(l)} = [\mathbf{U}_1^{(l)}; \dots; \mathbf{U}_P^{(l)}  ]$ with $\mathbf{U}_p^{(l)} = \mathbf{U}_{K,p}^{(l)} \odot \mathbf{U}_{R,p}^{(l)} \odot \mathbf{U}_{U,p}^{(l)}$, $\forall p \in \mathcal{P}$, and $\mathbf{U}_{K,p}^{(l)} = [\mathbf{u}_{K,p}({k^{(1)}}), \dots, \mathbf{u}_{K,p}({k^{(l)}})]$ (similarly for $\mathbf{U}_{R,p}^{(l)}$ and $\mathbf{U}_{U,p}^{(l)}$), the estimated cascaded channel $\dot{\mathbf{z}}$ consisting of active paths and the residual $\bm{\mathcal{T}}_p^{(l)}$ of the $l$-th iteration are given by $\dot{\mathbf{z}} \!=\! (\mathbf{U}^{(l)})^\dagger \mathbf{y}$ and $\bm{\mathcal{T}}_p^{(l)} \!=\! \bm{\mathcal{T}}_p^{(l-1)} \!- \! [\dot{\mathbf{z}}]_l \circ \mathbf{u}_{U,p}({u^{(l)}}) \circ u_{R,p}({m^{(l)}}) \circ \mathbf{u}_{K,p}({k^{(l)}})$, respectively. The above scanning procedure is iteratively executed until the residual power $\| \bm{\mathcal{T}}_p^{(l)} \|_F^2$   is below a predefined threshold $\varepsilon_{\text{I}}$. After convergence, $\dot{\mathcal{N}}$ stores all the linear indices $n^{(l)}$ of the identified active paths with $n^{(l)} = (k^{(l)}-1)\bar{N}_R N_U + (m^{(l)}-1)N_U + u^{(l)}$, $\forall l$,  and the estimated channel magnitudes $\{\dot{\nu}_{n^{(l)}}\}$ and phases  $\{ \dot{\omega}_{n^{(l)}} \}$ can be directly obtained from $\dot{\mathbf{z}}$.

\par On the other hand, for given $\dot{\mathcal{N}}$ and  $\dot{\mathbf{z}}$,  the DFO $f_d$ and off-grid vectors $\bm{\Xi}$ are estimated by solving the following ML problem:
\begin{align}
	\label{ML for DFO and off grids}
	\min_{f_d, \bm{\Xi} } \; \sum_{k=\!1}^{\bar{K}} \sum_{p=\!1}^{P}\! \| \mathbf{y}_{\!k,p} \!\!-\!\! \sqrt{p_T}x_{k,p}\mathbf{W}_p^{\!H} \mathbf{a}_{\!B}(\vartheta_{\!B}) \mathbf{f}_p^T \dot{\mathbf{R}}_{k,p} \mathbf{v}_p   \|^2,
\end{align}
where $\dot{\mathbf{R}}_{k,p} \!=\! \sum_{l} [\dot{\mathbf{z}}]_l e^{-j 2\pi f_k (\tau_{k^{(l)}} + \Delta \tau_{k^{(l)}})} \tilde{\mathbf{a}}_{U,p} (\varphi_{u^{(l)}} + \Delta \varphi_{u^{(l)}}) \\ 
\mathbf{a}_{\!R}^{\!T}(\! \bar{\vartheta}_{\!m^{\!(l)}} \!+\! \Delta\! \bar{\vartheta}_{\!m^{\!(l)}}\!, \bar{r}_{\!m^{\!(l)}} \!\!+\!\! \Delta\! \bar{r}_{\!m^{\!(l)}} \! ) $ and $\tilde{\mathbf{a}}_{U\!,p} \!(\varphi)  \!\triangleq\! \mathbf{a}_{\!U}^*\! (\varphi)  e^{j 2\pi \! f_d (p\!-\!1\!)T_{\!s} \varphi} $.
The ML estimates of $\dot{f}_{\!d}$ and $\{\dot{\Xi}_{\!u^{\!(l)}\!\!, m^{\!(l)}\!\!, k^{\!(l)}}\!\}$ can be derived using the gradient descent method as
\begin{align}
	\label{gradient descent ML}
	f_d^{(i\!+\!1)} \!\!=\!\! f_d^{(i)} \!-\! \epsilon_{f_d}^{(i)} g_{f_d}^{(i)}, \quad  
	\Delta \varphi_{u^{\!(l)}}^{(i\!+\!1)} \!\!=\!\! \Delta \varphi_{u^{\!(l)}}^{(i)} \!-\! \epsilon_{\varphi}^{(i)} g_{\varphi,l}^{(i)}, \; \forall l,
\end{align}
where $i$ denotes the iteration number, $\epsilon_{f_d}^{(i)}$ and $\epsilon_{\varphi}^{(i)}$
are the step sizes determined by the Armijo rule, $g_{f_d}^{(i)}$ and $g_{\varphi,l}^{(i)}$ are the gradients of the objective function in \eqref{ML for DFO and off grids} with respect to (w.r.t.) $f_d$ and $\Delta \varphi_{u^{\!(l)}}$, respectively. The updates for $ \Delta \bar{\vartheta}_{m^{\!(l)}}$, $\Delta \bar{r}_{m^{\!(l)}}$ and $\Delta \tau_{k^{(l)}}$ are conducted in a similar manner $\Delta \varphi_{u^{\!(l)}}$.

\subsection{Stage II: T-SPVBI Method}
\label{subsec: T-SPVBI}

\par Since straightforward approximation of the joint distribution, as in \eqref{approx joint dirstb}, fails to characterize the spatially structured sparsity in the CE phase, we combine the message passing and SPVBI methods via the turbo framework, referred to as the turbo-SPVBI (T-SPVBI) algorithm, as depicted in Fig. \ref{pic:T_SPVBI}, to effectively capture both the 2D block sparsity in the polar domain and the 1D clustered sparsity in the angle and delay domains. Particularly, Module A is an SPVBI estimator, processing inputs of the initial observation $\mathbf{y}^{(1)}$, coarse estimates from stage I and messages from Module B, whereas Module B acts as a spatially structured sparsity combiner, performing message passing over the MRF and Markov chains as discussed in Section \ref{subsec: spa-struct-sps model for CE} based on extrinsic messages from Module A. These modules are executed iteratively until convergence. It is noteworthy that in the CT phase, Module B is removed and the approximate marginal posteriors from the previous frame become the prior information for Module A. Detailed descriptions of these modules are provided below, with superscript $(t)$ omitted for clarity.
 
\subsubsection{Module A}
For convenience, let $\bm{\mathcal{H}}_1$ and $\bm{\mathcal{H}}_2$ denote two subsets of $\bm{\kappa}$, where $\bm{\mathcal{H}}_1 \triangleq \{  \bm{\nu}, \bm{\omega}, f_d, \Delta \bm{\varphi}, \Delta \bar{\bm{\vartheta}}, \Delta \bar{\mathbf{r}}, \Delta \bm{\tau} \}$ and $\bm{\mathcal{H}}_2 \triangleq \{ \mathbf{s}, \mathbf{s}_U, \mathbf{s}_R, \mathbf{s}_K\}$ for the CE phase; $\bm{\mathcal{H}}_2 \triangleq \{ \mathbf{s}\}$ for the CT phase. Furthermore, let  $\bm{\kappa}^l$ denote an individual variable in $\bm{\kappa}$ and $\kappa_n^l$ the $n$-th element of $\bm{\kappa}^l$, where $n \in \dot{\mathcal{N}}$ for $\bm{\kappa}^l \in \{ \bm{\nu}, \bm{\omega}, \mathbf{s} \}$; $n \in \dot{\mathcal{U}}$ for $\bm{\kappa}^l \in \{ \mathbf{s}_U, \Delta \bm{\varphi} \}$; $n \in \dot{\mathcal{M}}$ for $\bm{\kappa}^l \in \{ \mathbf{s}_R, \Delta \bar{\bm{\vartheta}}, \Delta \bar{\mathbf{r}} \}$; and $n \in \dot{\mathcal{K}}$ for $\bm{\kappa}^l \in \{ \mathbf{s}_K, \Delta \bm{\tau} \}$, with sets $\dot{\mathcal{U}}$, $ \dot{\mathcal{M}}$ and $\dot{\mathcal{K}}$ including the 3D indices derived from the linear indices in $\dot{\mathcal{N}}$. The SPVBI method employs the discrete variational distribution $q(\bm{\kappa}) \overset{(a)}{=} \prod_{\bm{\kappa}^{l} \in \bm{\kappa} } \prod_{n} q(\kappa_n^l) $ to approximate the posterior distribution $\hat{p}( \bm{\kappa} | \mathbf{y}) \propto \hat{p}(\mathbf{y}, \bm{\kappa}) $, where $(a)$ is due to the mean-field assumption \cite{parisi1988statistical}, $ q(\kappa_n^l) = \sum_{k = 1}^{N_p} w_{n,k}^l \delta(\kappa_n^l- p_{n,k}^l)$ with $N_p$ being the number of particles, $\mathbf{w}_n^l = [w_{n,1}^l, \dots, w_{n,N_p}^l]^T$ and $\mathbf{p}_n^l = [p_{n,1}^l, \dots, p_{n,N_p}^l]^T$ being the weights and positions of the particles. Respectively, $\hat{p}(\mathbf{y}, \bm{\kappa})$ is specified in \eqref{approx joint dirstb} for the CT phase, and for the CE phase, $\hat{p}(\mathbf{y}, \bm{\kappa})$ is derived by replacing the original priors $p(\mathbf{s}_U)$, $p(\mathbf{s}_R)$ and $p(\mathbf{s}_K)$ in $p(\mathbf{y}, \bm{\kappa})$  by $\hat{p}(\mathbf{s}_U)$, $\hat{p}(\mathbf{s}_R)$ and $\hat{p}(\mathbf{s}_K)$, respectively, where $\hat{p}(\mathbf{s}_U) = \prod_{n \in \dot{\mathcal{U}}} \pi_{U,n}^{\mathbb{I}(s_{U,n})}
(1-\pi_{U,n})^{(1-\mathbb{I}(s_{U,n})) }$ with $\pi_{U,n} = g_{U,n}^A(1)/(g_{U,n}^A(-\!1) + g_{U,n}^A(1))$ and $\{g_{U,n}^A(s_{U,n})\}$ denoting the output messages from Module B,  $\hat{p}(\mathbf{s}_R)$  and $\hat{p}(\mathbf{s}_K)$ are defined similarly. Note that for $q(\kappa_n^l)$, $\forall \bm{\kappa}^l \in \bm{\mathcal{H}}_2$, $\forall n$,  the  number of  particles  is set to $N_p \!=\! 2$ with their positions fixed at $p^l_{\!n,1} \!\!=\!\! 1$ and $p^l_{\!n,2} \!\!=\!\! -\!1$. Subsequently, the particle weights $\{\mathbf{w}_n^l\}$ and positions $\{\mathbf{p}_n^l\}$ are optimized to minimize the Kullback-Leibler (KL) divergence between $q(\bm{\kappa})$ and $\hat{p}(\mathbf{y}, \bm{\kappa})$, denoted by $K\big( \{\mathbf{w}_n^l\}, \{  \mathbf{p}_n^l \} \big)$,  with the optimization problem formulated as
\begin{subequations}
	\label{KLD min prob}
	\begin{align}
		\!\!\!\min_{ \{\mathbf{w}_n^l\}, \{  \mathbf{p}_n^l \}  } 
		& \;\; K\big( \{\mathbf{w}_n^l\}, \{  \mathbf{p}_n^l \} \big) \triangleq \int q(\bm{\kappa})\ln \frac{q(\bm{\kappa})}{\hat{p}(\mathbf{y}, \bm{\kappa})} d\bm{\kappa} \label{KLD min prob, obj func} \\
		\text{s.t.} & \;\;   \sum_{k=1}^{N_p} \! w^l_{n,k} \!\!=\! 1, \;  \epsilon_w \!<\! w^l_{n,k} \!\leq\! 1, \; \forall n,k, \forall \bm{\kappa}^l \!\in\! \bm{\kappa},  \label{sum_1 constant, w} \\
		& \;\;  \dot{\kappa}_n^l \!-\! \Delta_n^l \!\leq\! p^l_{n,k} \!\leq\! \dot{\kappa}_n^l \!+\! \Delta_n^l, \; \forall n,k, \forall \bm{\kappa}^l \!\in\! \bm{\mathcal{H}}_1,   \label{position range constraint, p}
		\vspace{-0.0cm}
	\end{align}
\end{subequations}
where $\epsilon_w$ is set to a small positive value to prevent the occurrence of meaningless particles, $\{\dot{\kappa}_n^l\}$ represent the coarse estimates from stage I, and $\{\Delta_n^l\}$ denote the estimation error ranges associated with $\{\dot{\kappa}_n^l\}$. Since the truth value of $\kappa_n^l$ most likely lies in the interval  $[\dot{\kappa}_n^l - \Delta_n^l, \dot{\kappa}_n^l + \Delta_n^l]$,  constraints \eqref{position range constraint, p} are imposed to enhance the convergence rate.\footnote{Estimation error ranges $\{\Delta_n^l\}$ for $\{\dot{\nu}_n\}$ and $\{\dot{\omega}_n\}$ can be derived from the statistical error analysis of the OMP algorithm \cite{tan2018omp}, whereas determining those for $\dot{f}_d$ and $\{\dot{\Xi}_{u,m,k}\}$ is challenging due to the application of the gradient descent method. In simulations, we set $\Delta_n^l = c |\dot{\kappa}_n^l| $ for $\Delta_n^l$ related to $\dot{f}_d$ and $\{\dot{\Xi}_{u,m,k}\}$, with $c$ being a scaling factor that varies from 0 to 1.} 
Problem \eqref{KLD min prob} is challenging to solve because the objective function \eqref{KLD min prob, obj func} cannot be explicitly expressed due to the high-dimensional integration. 
To address this challenge, we employ the stochastic successive convex approximation (SSCA) method \cite{liu2019ssca} to alternately optimize the particle positions and weights. Detailed derivations for $\mathbf{w}_n^l$ and $\mathbf{p}_n^l$ are provided in Appendix B. Let $\hat{\mathbf{w}}_{n}^{l}$ and $\hat{\mathbf{p}}_{n}^{l}$ denote the converged particle weights and positions of $q(\kappa_n^{l})$, $\forall \bm{\kappa}^l \in \bm{\mathcal{H}}_1$, $\forall n$. Then, the extrinsic messages w.r.t. $\mathbf{s}_{U}$ from Module A, denoted by $\{ g_{U,n}^B(s_{U,n}) \}$, are calculated as $g_{U,n}^B(s_{U,n}) = q(s_{U,n})/g_{U,n}^A(s_{U,n})$, $\forall n$. The extrinsic messages w.r.t. $\mathbf{s}_{R}$ and $\mathbf{s}_{K}$ are derived in a similar manner. Besides, the mean and variance of the Gaussian approximation $\hat{q}(\kappa_n^{l}) = \mathcal{N}(\hat{m}_n^{l}, \hat{\Sigma}_n^{l})$ of $q(\kappa_n^{l})$, as required in \eqref{approx joint dirstb}, are calculated as
$\hat{m}_n^{l} = \mathbb{E}_{q(\kappa_n^{l})}\big[\kappa_n^{l}\big] \!=\!\sum_{k=1}^{N_p} \hat{p}_{n,k}^{l} \hat{w}_{n,k}^{l}$ and
$\hat{\Sigma}_n^{l} \!=\! \mathbb{E}_{q(\kappa_n^{l})}\big[\big(\kappa_n^{l} \!- \hat{m}_n^{l}\big)^{\!2}\big] \!=\! \sum_{k\!=\!1}^{N_p}\!\big(\hat{p}_{n,k}^{l}\big)^{\!2}  \hat{w}_{n,k}^{l} \!-\! \big(\hat{m}_n^{l}\big)^{\!2}$.

\subsubsection{Module B}
The factor graphs of $\mathbf{s}_R$ and $\mathbf{s}_U$ are illustrated in Fig. \ref{pic:MC}(\subref{pic:MRF_angle_MC}), with the factor graph of $\mathbf{s}_K$ resembling that of $\mathbf{s}_U$. Note that the extrinsic messages from Module A, i.e., $\{ g_{U\!,u}^B\!(s_{U\!,u}), g_{R,m}^B\!(s_{\!R,m}), g_{K\!,k}^B\!(s_{U,k}) \}$, $\forall u \in \dot{\mathcal{U}}$, $\forall m \in \dot{\mathcal{M}}$,  $\forall k \in \dot{\mathcal{K}}$, act as the prior factor nodes within their respective factor graphs. By performing sum-product message passing over these graphs, all the messages are effectively obtained. Importantly, due to the presence of loops within the MRF, the loopy belief propagation method \cite{li2009markov} is employed for message updates.
Upon completing the update of all the messages, the output messages from Module B are given by  $\{ g_{U,u}^A(s_{U,u}) = \eta_{s_{U,u}\to g_{U,u}^B}(s_{U,u}) \}$, $\{ g_{R,m}^A(s_{R,m}) = \eta_{s_{R,m}\to g_{R,m}^B}(s_{R,m}) \}$ and $\{ g_{K,k}^A(s_{K,k}) = \eta_{s_{K,k}\to g_{K,k}^B}(s_{K,k}) \}$.

The overall complexity of the proposed TS-CET algorithm can be expressed as $\mathcal{O}\big(N_{\text{I}}(N_{JS} R_B \bar{K} L^3 + N_U^2 + \bar{N}_R^2 + N_f^2) + 2J N_{\text{II}} N_{SSCA}(N_p B F_{\text{grad}} + N_p^3)\big)$, where the derivation details and symbol definitions are provided in Appendix C for clarity.

\renewcommand\baselinestretch{\linspreadalgr}\selectfont 
\begin{algorithm}[htb]
	\caption{Proposed TS-CET algorithm}
	\label{Algorithm1, TS-CET algorithm}
	\textbf{Input}: $\mathbf{y}^{\!(t)}$\!, $\mathbf{W}_p^{\!(t)}$\!, $\mathbf{f}_p^{\!(t)}$\!, $\mathbf{v}_p^{\!(t)}$, $\mathbf{x}_{\!p}$, $\tilde{\mathbf{A}}_{\!B\!U\!,p}(\mathbf{0})$, $\forall p$, $\mathbf{A}_{\!R}(\mathbf{0},\!\mathbf{0})$, $\mathbf{A}_{f}(\mathbf{0})$, 
	$q(\!\kappa_n^{\!l,(t\!-\!1)}\!)$ in the CT phase, $\forall  \bm{\kappa}^{l}$, $\forall n$, 
	maximum iteration numbers $N_{\text{I}}$, $N_{\text{II}}$, iteration indices $i_{\text{I}}$, $i_{\text{II}}=1$. \\
	\textbf{Output}: MMSE estimates $\hat{m}_n^{l,(t)}$, $\forall  \bm{\kappa}^{l} \in \bm{\mathcal{H}}_1$, $\forall n$. \!\!\!\!\!\!\!\!\!\!\!\!\!\!\!\!\!\!\!\!\!\!\!\!\!\!\!\!\!\!\!\!\!\!\!\!\!\!\!\!\!\!\!\!\! 
	\begin{algorithmic}[1]
        \STATE {\textbf{\%Stage I: ETJM-OMP Method}}
        \STATE {Initialize coarse 3D index $\mathcal{S}^{(t)}$. }
        \STATE { \textbf{For} $i_{\text{I}} = 1, \dots, N_{\text{I}}$}
		\STATE {\; \textbf{While} not converged \textbf{do} }\!\!\!
		\STATE {\;\;\;\; Find 3D indices of active paths using \eqref{ETJM-OMP} in $\mathcal{S}^{(t)}$.}
		\STATE {\;\;\;\; Calculate $\dot{\mathbf{z}}^{(t)}$ and update the linear index set $\dot{\mathcal{N}}^{(t)}$.}
		\STATE {\; \textbf{end While} }
		\STATE {\; Update $\dot{f}_d^{(t)}$ and  $\{\dot{\Xi}_{u,m,k}^{(t)}\}$ according to \eqref{gradient descent ML}.\!\!\!\!\!\!\!\!\!\!\!\!\!\!\!}
		\STATE {\textbf{end For} }
        \STATE {\textbf{\%Stage II: T-SPVBI Method}}
        \STATE {\textbf{For} $i_{\text{II}} = 1, \dots, N_{\text{II}}$}
		\STATE {\; \textbf{\%Module A: SPVBI Estimator}}  
		\STATE {\; Initialize $\{\mathbf{w}_n^{l,(t)}\}$ and $\{\mathbf{p}_n^{l,(t)}\}$, $\forall \bm{\kappa}^{l,(t)} \in \bm{\kappa}^{(t)}, \forall n$.  } 
		\STATE {\; \textbf{While} not convergence \textbf{do} }
		\STATE {\;\;\;\; \textbf{For} $\kappa_n^{l,(t)} \in \bm{\kappa}^{(t)}$}
        \STATE {\;\;\;\; \;\;  Update $\mathbf{p}_n^{\!l\!,(t)}\!$ and $\mathbf{w}_n^{l\!,(t)}\!$ by solving \eqref{KLD min prob} with SSCA.}
		\STATE {\;\;\;\; \textbf{end For} }
		\STATE {\; \textbf{end While} }
		\STATE {\;\textbf{\%Module B: Sparsity Combiner} (for CE phase) }  
		\STATE {\; Calculate $\{ g_{U\!,u}^B\!(s_{U\!,u}), g_{R,m}^B\!(s_{\!R,m}), g_{K\!,k}^B\!(s_{U,k}) \}$, $\forall u,m,k$.}\!\!\!
		\STATE {\; Perform sum-product message passing over the  support \\  \;  factor graphs  and calculate the output messages. }
		\STATE {\textbf{end For} }
	\end{algorithmic}
\end{algorithm}
\vspace{-0.1cm}
\renewcommand\baselinestretch{\linspread}\selectfont

\section{Simulation results}
\label{sec:simulation}
In this section, we provide numerical results to validate the effectiveness of the proposed TS-CET algorithm. In our simulations, a 2D coordinate system is considered, with a fixed BS and XL-IRS placed at $[0 \, \text{m}, 0 \, \text{m}]$ and $[60 \, \text{m}, 30 \, \text{m}]$, respectively. 
The initial position of the mobile user is $[70 \, \text{m}, 5 \, \text{m}]$, and the user moves along the y-axis at a speed of $v = 5 \, \text{m/s}$. 
The BS is equipped with $N_B = 32$ antennas and $R_B = 4$ RF chains, while the XL-IRS has $N_R = 128$ reflecting elements. 
The user is equipped with $N_U = 16$ antennas and one RF chain. The polar domain's non-uniform position grid $[\bar{\boldsymbol{\theta}}, \mathbf{\bar{r}}]$ contains $\overline{N}_R = 384$ grid points, 
with $N_R = 128$ angle grid points and $S = 3$ distance grid points. The IRS-user channel consists of three block targets, which serve as scatterers for the initial frame (i.e., $t = 1$). 
These block targets correspond to 4, 2, and 2 scatterer paths, respectively, forming $L^{(1)} = 8$ paths. They also induce 2D block sparsity in the polar domain and 1D cluster sparsity in both the angle and delay domains of the sparse cascade channel $\bm{\mathcal{Z}}^{(1)}$.
The system bandwidth is $f_s = 200$ MHz with $k = 128$ subcarriers. The cyclic prefix (CP) length is set to $L_{cp} = k/4$, which yields $N_f = 3L_{cp}/4 = 24$ delay grid points that cover the effective channel delay spread. For the CET process, only $K = 31$ subcarriers are utilized.
In the CE phase, $P_{\text{I}}= 20$ OFDM pilot symbols are used, while the CT phase uses $P_{\text{II}} = 6$ OFDM pilot symbols. 
The large-scale path loss of the channel is modeled according to the method in  \cite{chen2023path}. 
Unless otherwise stated, other system parameters are as follows: carrier frequency 28 GHz, $\sigma^2 = -91$ dBm, SNR = 10 dB, $\varepsilon_v = 0.5$, $N_p = 10$, $B = 15$. Based on the above parameters, the Rayleigh distance of the XL-IRS and the user’s distance to the XL-IRS are approximately 86.2 m and 26.9 m, respectively, which indicates that the user is located in the near-field region of the XL-IRS.
In this paper, the normalized mean squared error (NMSE) of the channel $\bm{\mathcal{R}}_p^{(t)}$ is chosen as the performance metric, which is defined as $\frac{1}{P} \sum_{p=1}^{P} \frac{\left\|\bm{\mathcal{\hat{R}}}_p^{(t)} - \bm{\mathcal{R}}_p^{(t)}\right\|_F^2}{\|\bm{\mathcal{R}}_p^{(t)}\|_F^2},$ where $\bm{\mathcal{\hat{R}}}_p^{(t)}$ denotes the reconstructed cascaded channel estimate as defined in \eqref{eq:sparse repre of cas_chan}.
Then for performance comparison, we consider the following baseline algorithms:

\begin{itemize}
    \item \textbf{EM-OMP for the CE phase}: This algorithm adopts the EM framework and employs the OMP algorithm in the E-step to estimate the sparse cascaded channel, meanwhile a solvable surrogate function is constructed in the M-step to estimate the DFO and off-grid vectors.
   
    \item \textbf{EM-FSBL for the CE phase}: Similar to EM-OMP, this algorithm replaces the OMP algorithm in the E-step with the FSBL algorithm.

    \item \textbf{EM-FSBL for the CT phase without Channel Tracking}: This algorithm does not exploit the temporal correlation of the channel across different frames. Instead, it independently applies the EM-FSBL algorithm to each frame to estimate the cascaded channel.

    \item \textbf{TS-CET for the CT phase without Non-Ideal Parameter Tracking}: This benchmark employs the proposed TS-CET algorithm, but only  estimates the non-ideal parameters $\mathbf{\Xi}$ and $f_d$ in the CE phase. In the CT phase, historical non-ideal parameter estimates are used, and only the sparse cascade channel gains are estimated.
    
    \item \textbf{TS-CET with Perfect Non-Ideal Parameters}: This version of the proposed TS-CET algorithm assumes perfect knowledge of non-ideal parameters, serving as a performance lower bound for channel estimation accuracy.
\end{itemize}

\begin{figure}[tbp]
	\setlength{\abovecaptionskip}{-0.0cm}
	\centering
	{\includegraphics[width=0.4\textwidth]{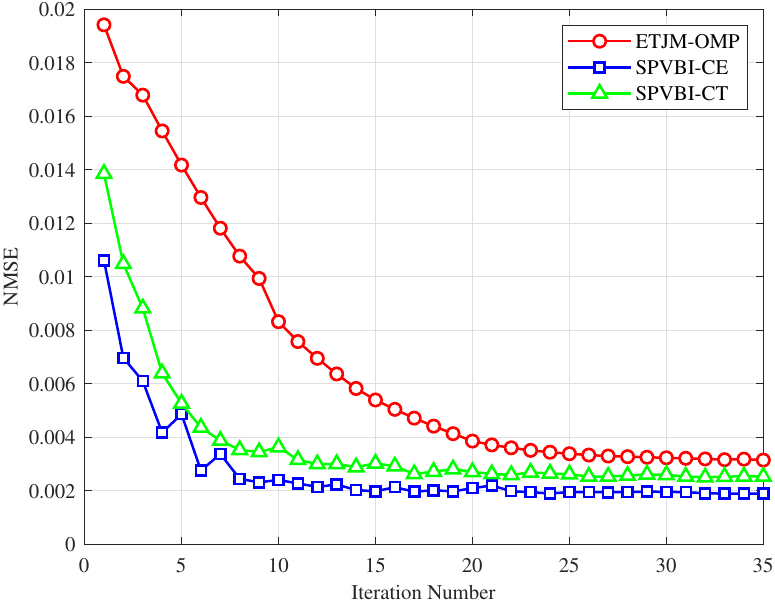}}
	\caption{Convergence behavior of the  ETJM-OMP and SPVBI algorithms.}
	\label{pic:Convergence performance}
	\vspace{-0.3cm}
\end{figure}

Before comparing the performance results, the convergence behaviors of the ETJM-OMP algorithm and SPVBI algorithms used in both phases are first verified in Fig. \ref{pic:Convergence performance}.
As illustrated, the ETJM-OMP algorithm converges after approximately 30 iterations, and its convergence curve exhibits a monotonically decreasing trend. Moreover, the SPVBI algorithm used in the CE and CT phases converge after around 25 iterations.
Their convergence curves show slight oscillations during the early iterations but gradually decrease in a nearly monotonic manner as the iterations proceed.
This phenomenon arises because the proposed SPVBI method is developed based on the SSCA framework, which recursively approximates the true gradient through statistical expectations of the generated sample gradients.
As such, minor oscillations during the early iterations are expected, since stochastic optimization algorithms generally do not guarantee strictly monotonic convergence.

\begin{figure}[tbp]
	\setlength{\abovecaptionskip}{-0.0cm}
	\centering
	{\includegraphics[width=0.4\textwidth]{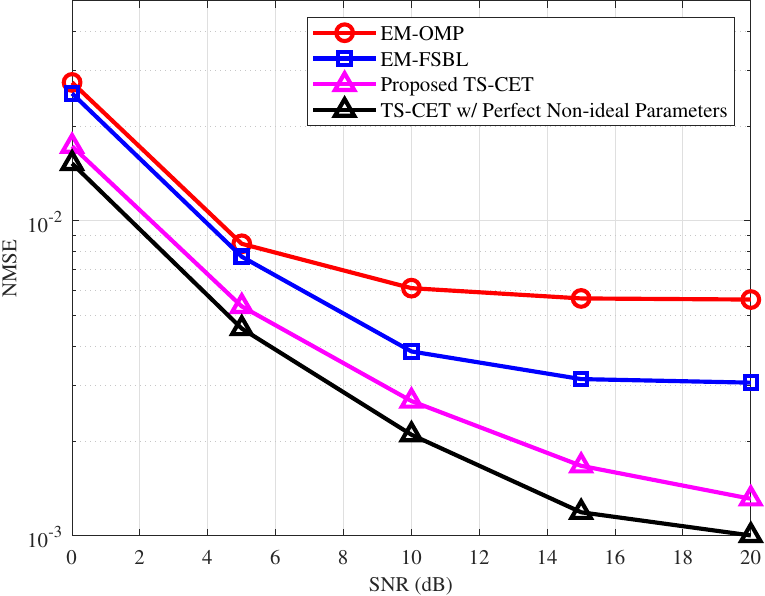}}
	\caption{NMSE performance of the considered algorithms versus SNR in the CE phase.}
	\label{pic:NMSE CE}
	\vspace{-0.3cm}
\end{figure}
Fig. \ref{pic:NMSE CE} investigates the NMSE performance achieved by the considered algorithms in the CE phase at different SNR levels.
As shown, the proposed TS-CET algorithm performs better than the EM-OMP and EM-FSBL algorithms , which is mainly due to the following two reasons: 
1) the proposed algorithm utilizes a particle-based posterior estimation method, which enables a more accurate complex double Gaussian prior for solving the considered sparse Bayesian CE problem; 
2) the proposed algorithm adopts a turbo framework that combines 2D block sparsity in the polar domain, 1D cluster sparsity in the angle and delay domains within Module B, which effectively reduces the false alarm probability of active paths, thus enhancing estimation performance. 
Additionally, it can be observed that while there is a small performance gap between the proposed TS-CET algorithm and its ideal version with perfect non-ideal parameters, this gap gradually stabilizes as the SNR increases. This behavior is mainly due to the sampling approximation step in SPVBI, which limits the reduction of parameter estimation errors.

\begin{figure}[tbp]
	\setlength{\abovecaptionskip}{-0.0cm}
	\centering
	{\includegraphics[width=0.4\textwidth]{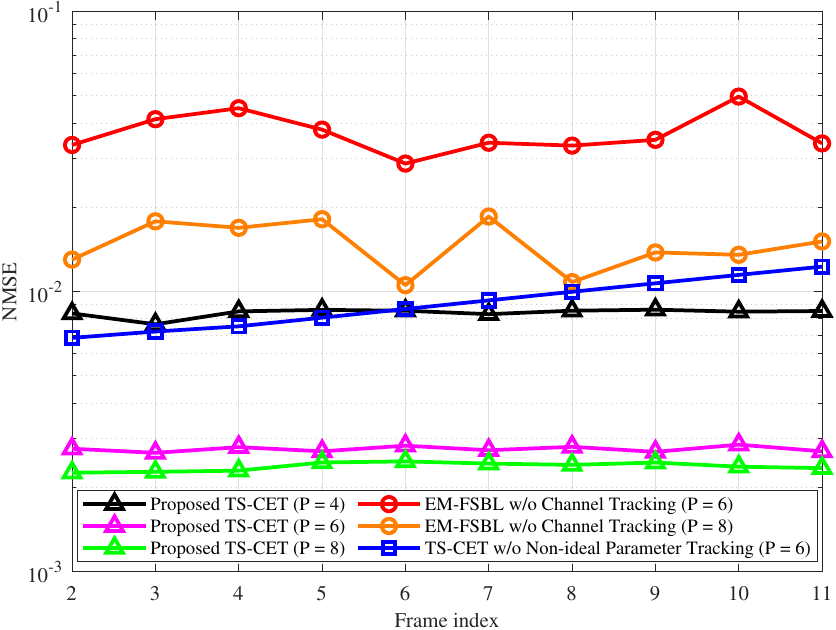}}
	\caption{NMSE performance under different pilot numbers and time frames in the CT phase.}
	\label{pic:NMSE CT vs different pilots}
	\vspace{-0.3cm}
\end{figure}

Fig. \ref{pic:NMSE CT vs different pilots} compares the NMSE performance in the CT phase over different time frames and with different tracking pilot numbers. 
It can be observed that the proposed TS-CET algorithm still outperforms the baseline methods, achieving stable channel tracking. In contrast, the EM-FSBL algorithm does not exploit the temporal correlation of the channels and performs complete channel estimation independently in each frame. Consequently, its performance fluctuates across frames, and the NMSE significantly degrades when the number of pilots in the CT phase is substantially reduced.
Furthermore, when the TS-CET algorithm does not track the non-ideal parameters during the CT phase, its performance degrades significantly, as evidenced by a steady rise in NMSE over time due to the accumulation of estimation errors across frames.
On the other hand, by examining the results under different pilot numbers, it is observed that the proposed TS-CET algorithm achieves excellent channel tracking performance even with only 6 pilots. Increasing the number of pilots brings little further improvement, while reducing it to 4 leads to a clear degradation in performance. Therefore, 6 pilots are sufficient for reliable channel tracking in the CT phase. Meanwhile, the EM-FSBL algorithm shows a remarkable NMSE improvement when the pilot number increases to 8, implying that pilot number acts as a key performance bottleneck for this algorithm.

\begin{figure}[tbp]
	\setlength{\abovecaptionskip}{-0.0cm}
	\centering
	{\includegraphics[width=0.4\textwidth]{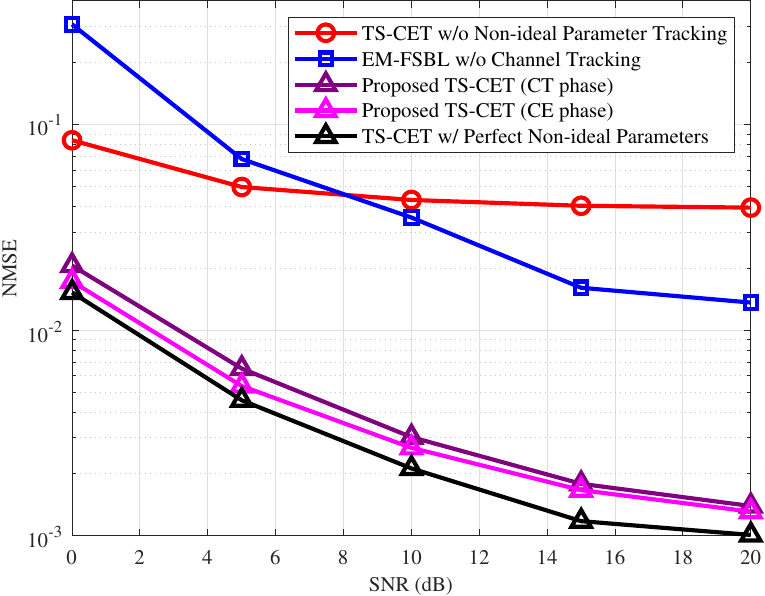}}
	\caption{NMSE performance of the considered algorithms versus SNR in the CT phase.}
	\label{pic:NMSE CT}
	\vspace{-0.3cm}
\end{figure}
Finally, Fig. \ref{pic:NMSE CT} presents the time-averaged NMSE (TNMSE) of the CT phase under different SNR conditions, where TNMSE is defined as $ \frac{1}{(T-1)P}\sum_{t=2}^{T}\sum_{p=1}^{P} \frac{\left\|\bm{\mathcal{\hat{R}}}_p^{(t)} - \bm{\mathcal{R}}_p^{(t)}\right\|_F^2}{\|\bm{\mathcal{R}}_p^{(t)}\|_F^2}$, with $T = 50$ time frames. 
It can be observed that as the SNR increases, the TNMSE performance of all algorithms steadily improves. 
However, the EM-FSBL algorithm is limited by the number of pilots, and the TS-CET algorithm has inherent errors when non-ideal parameters are not tracked, both of which fail to achieve good channel tracking performance, 
resulting in a significant performance gap compared to the proposed TS-CET algorithm. Further observation reveals that the performance of the proposed TS-CET algorithm in the CT phase is close to that in the CE phase, demonstrating that the proposed algorithm can maintain nearly identical estimation performance in the CT phase, even with a reduced number of pilots. This close alignment in performance between the two phases is especially pronounced at high SNR. In such cases, the effect of noise becomes less dominant, allowing the channel’s temporal correlation to be better exploited.

\section{Conclusions}
\label{sec:conclusion}
In this paper, we addressed the near-field sparse CET problem in XL-IRS-aided wideband mmWave systems. 
To exploit the inherent multi-dimensional sparsity of the cascaded BS-IRS-user channel, we constructed a tensor-based sparse representation in the polar-angle-delay domain by designing a sparse dictionary and leveraging tensor representation techniques.
Based on this representation, spatially structured and temporally dynamic sparse priors were designed for the CE and CT phases, respectively.
Building upon these priors, we developed the TS-CET algorithm, which integrates tensor-based OMP, particle-based VBI, and message passing to efficiently approximate the posterior distributions of channel parameters in both phases while achieving low pilot and computational overhead.
Simulation results demonstrated that the proposed TS-CET algorithm achieves high estimation accuracy in both CE and CT phases, with low pilot overhead.

\section*{Appendix A \\  Passive Beamforming Design for XL-IRS}
\label{Passive Beamforming Design in CT phase}
As mentioned in the main text, we adopt different design strategies for the passive beamforming vectors $\mathbf{v}_p^{(t)}$ at the XL-IRS in different phases.
Specifically, in phase I, since there is no prior information about the user and scatterer positions, $\mathbf{v}_p^{(1)}$ is designed using the following random phase strategy, i.e.,
\begin{align}
	\label{bf design for v}
	[\mathbf{v}_p^{(1)}]_n = e^{j\theta_{n,p}^v},  \; \forall n \in \mathcal{N}_R, \; p = 1, \!\cdots\!, P,
\end{align}
where $\theta_{n,p}^v$ is randomly generated from $[0,2\pi]$.
Adopting this random phase design for $\mathbf{v}_p^{(1)}$ in phase I offers two advantages: 1) omnidirectional beams at the XL-IRS are constructed to detect all possible user and scatterer positions, and 2) the correlation between different columns of the measurement matrix $\mathbf{F}^{(1)}$ is reduced, thus enhancing the performance of the CS algorithm. In phase II, the primary objective for designing $\mathbf{v}_p^{\!(t)}$ is to strike a balance between 1) exploiting the candidate positions of users and scatterers (obtained from the previous frame) to achieve the passive beamforming gain at the XL-IRS, and 2) exploring unknown positions to detect potential user and/or scatterers. 
To this end, we propose a near-field multi-beam design method for $\mathbf{v}_p^{\!(t)}$ in this phase. In particular, let the set $\bar{\mathcal{N}}_R^{(t-1)}$ include all the non-zero element indices of the estimated position support vector $\hat{\mathbf{s}}_{R}^{(t-1)}$ in the $(t\!-\!1)$-th frame.\footnote{Note that $\hat{\mathbf{s}}_{R}^{(t-1)}$ can be directly obtained from the CE result for the case of $t=2$, while for $t>2$, $\hat{\mathbf{s}}_{R}^{(t-1)}$ is derived based on its relationship with the estimated support vector $\hat{\mathbf{s}}^{(t-1)}$.}
Based on the previous support set, we construct an expanded candidate set $\widetilde{\mathcal{N}}_R^{(t)}$. Specifically, $\widetilde{\mathcal{N}}_R^{(t)}$ is composed of $\bar{\mathcal{N}}_R^{(t-1)}$ and its neighboring indices in the distance–angle domain, where the neighborhood size is controlled by $\Delta N$, ensuring that potential channel variations can be effectively captured.
Then, we define $\tilde{\mathbf{A}}_R^{(t)} \!=\! [\mathbf{a}_R(\bar{\vartheta}_{n_1} , \bar{r}_{n_1}), \!\cdots\!, \mathbf{a}_R(\bar{\vartheta}_{n_{\tilde{Q}}}, \bar{r}_{n_{\tilde{Q}}})  )]$ as a matrix comprising $\tilde{Q} = |\tilde{\mathcal{N}}_R^{(t)}|$ near-field beams focused on the most potential positions with $n_q$ being the $q$-th ($q= 1,\cdots, \tilde{Q}$) element of $\tilde{\mathcal{N}}_R^{(t)}$. 
To explore the remaining $\dot{Q} \!=\! \bar{N}_R \!-\! \tilde{Q}$ positions, identified by the index set $\dot{\mathcal{N}}_R^{\!(t)} = \bar{\mathcal{N}}_R \backslash \tilde{\mathcal{N}}_R^{(t)}$,  in the position grid $[\bar{\bm{\vartheta}}, \bar{\mathbf{r}}]$, we introduce another matrix $\dot{\mathbf{A}}_{R}^{{\!(t)}} \!=\! [\mathbf{a}_R(\bar{\vartheta}_{m_1}, \bar{r}_{m_1} ), \!\cdots\!, \mathbf{a}_R(\bar{\vartheta}_{m_{\dot{Q}}}, \bar{r}_{m_{\dot{Q}}} )  )]$ consisting of $\dot{Q}$ near-field beams focused on remaining positions with $m_q$ being the $q$-th ($q= 1,\cdots, \dot{Q}$) element of $\dot{\mathcal{N}}_R^{\!(t)}$.

Then in the $p$-th pilot duration, our goal is  to design  $\mathbf{v}_p^{\!(t)}$ with multiple beams simultaneously focusing on the $\tilde{Q}$ positions of interest and a subset of $\dot{Q}$ remaining positions. By defining the aggregated matrix $\mathbf{A}_{R,p}^{{\!(t)}} = [\tilde{\mathbf{A}}_R^{(t)},\dot{\mathbf{A}}_{R,p}^{{\!(t)}}]$ where $\dot{\mathbf{A}}_{R,p}^{{\!(t)}}$ incorporates the columns of $\dot{\mathbf{A}}_{R}^{{\!(t)}}$  from $ (p-1)M_R^{\!(t)} + 1$ to $pM_R^{\!(t)}$ for $p < P$, and from $ (P-1)M_R^{\!(t)} + 1$ to $\dot{Q}$ for $p = P$, with $M_R^{\!(t)} = \lfloor \dot{Q}/P \rfloor $.
Therefore, the corresponding optimization problem can be formulated as
\begin{subequations}
	\label{opt-prob: v design}
	\begin{align}
		\max_{\mathbf{v}_{p}^{\!(t)}} & \;\;  \| (\mathbf{v}_{p}^{\!(t)})^T {\mathbf{A}}_{R,p}^{{\!(t)}}\|^2  \label{v-opt-prob, obj func} \\
		\text{s.t.} & \;\; |(\!\mathbf{v}_{p}^{\!(t)})^{\!T} \mathbf{a}_q|^2 \!-\! | (\mathbf{v}_{p}^{\!(t)})^{\!T} \mathbf{a}_1|^2 < \varepsilon_{v}, \; q \!=\! 2, \!\cdots\!, \check{Q},  \label{v-opt-prob, qusi-const} \\
		& \;\;  |[\mathbf{v}_{p}^{\!(t)}]_n| = 1, \; \forall n \in \mathcal{N}_R   \label{v-opt-prob, uni-mod},
		\vspace{-0.0cm}
	\end{align}
\end{subequations}
where  $\mathbf{a}_q$ denotes the $q$-th column of ${\mathbf{A}}_{R,p}^{{\!(t)}}$, $\check{Q}$ is the total number of columns in ${\mathbf{A}}_{R,p}^{{\!(t)}}$, and $\varepsilon_{v}$ represents a threshold constant. Note that the constraints \eqref{v-opt-prob, qusi-const} are applied here to ensure that $|(\mathbf{v}_{p}^{\!(t)})^T {\mathbf{A}}_{R,p}^{{\!(t)}}|$ forms a quasi-constant magnitude vector, thereby achieving similar beam gains across the generated multiple beams. 
Problem  \eqref{opt-prob: v design} can be effectively solved using the semidefinite relaxation (SDR) method \cite{luo2010semidefinite} combined with rank-one rescaling\cite{Cao2017semidefinite}, or other non-variable-lifting methods \cite{Beck2010SCA}, the details are omitted here for brevity. Besides, the design of the active beamforming vector $\mathbf{f}_p^{\!(t)}$ follows a procedure similar to that of $\mathbf{v}_p^{\!(t)}$.

\section*{Appendix B \\  Particle Weight and Position Optimization via SSCA}
As mentioned in the main text, to solve the optimization problem in \eqref{KLD min prob}, we propose to alternatively optimize the particle weights and positions using the SSCA method \cite{liu2019ssca}. Specifically, in the $i$-th iteration ($i \geq 1$), the optimization subproblem for $\mathbf{p}_n^l$, $\forall \bm{\kappa}^l \in \bm{\mathcal{H}}_1$, $\forall n$, is given by
\begin{subequations}
	\label{KLD min subprob p}
	\begin{align}
		\!\!\!\min_{   \mathbf{p}_n^l  } 
		& \;\; K_{\!\kappa_n^l}^{\!(i)} ( \mathbf{w}_n^{l, (i\!-\!1)}\!\!, \mathbf{p}_n^l ) \!\triangleq\! \mathbb{E}_{\sim \kappa_n^l} \!\big[ g_{\kappa_n^l}^{(i)} ( \mathbf{w}_n^{l, (i\!-\!1)}\!, \mathbf{p}_n^l; \sim \! \kappa_n^l )   \big] \label{KLD min subprob p, obj func} \\
		\text{s.t.} & \;\;  \dot{\kappa}_n^l - \Delta_n^l \leq p^l_{n,k} \leq \dot{\kappa}_n^l + \Delta_n^l, \; \forall k,   \label{KLD min subprob p, costraint p}
		\vspace{-0.0cm}
	\end{align}
\end{subequations}
where $ g_{\kappa_n^l}^{(i)} ( \mathbf{w}_n^{l}, \mathbf{p}_n^l; \sim  \kappa_n^l ) = \sum_{k=1}^{N_p} w_{n,k}^l  \ln w_{n,k}^l -  \sum_{k=1}^{N_p}  w_{n,k}^l \\ \ln p(\mathbf{y}, \sim \kappa_n^l, \kappa_n^l ) |_{\kappa_n^l = p_{n,k}^l}$, $\mathbf{w}_n^{l, (i-1)}$ denotes particle weights of $q(\kappa_n^l)$ from the $(i-1)$-th iteration with $w_{n,k}^{l, (0)} = 1/N_p$, $\forall k$, and $\sim  \kappa_n^l$ represents all variables in $\bm{\kappa}$ except for $\kappa_n^l$. To solve problem \eqref{KLD min subprob p}, we first construct a convex surrogate function for $K_{\!\kappa_n^l}^{\!(i)} ( \mathbf{w}_n^{l, (i\!-\!1)}\!\!, \mathbf{p}_n^l )$ as follows:
\begin{align}
	\label{surog fuc, p}
	\!\bar{f}_{\!\mathbf{p}_{n}^l}^{(i)} (\mathbf{p}_{n}^l ) \!\triangleq\! \big( \mathbf{f}_{\mathbf{p}_n^l}^{(i)} \big)^{\!\!T}\! \big( \mathbf{p}_{n}^l \!-\! \mathbf{p}_{n}^{l, (i\!-\!1)} \big) \!+\! \Omega_{\mathbf{p}_{n}^l} \| \big( \mathbf{p}_{n}^l \!-\! \mathbf{p}_{n}^{l, (i\!-\!1)} \|^2, 
\end{align}
where $\mathbf{p}_{n}^{\!l, (i\!-\!1)}\!$ denotes the particle positions of $q(\!\kappa_n^l)$ from the $(i\!\!-\!\!1)$-th iteration with $p_{\!n,k}^{\!l, (0)} \!\!=\!\! \dot{\kappa}_n^l \!\!-\!\! \frac{2\Delta_n^l}{N_p}(\frac{N_p\!+\!1}{2}\!\!-\!\!k)$, $\forall k$, $\Omega_{\mathbf{p}_{n}^l}$ is a positive constant, and $\mathbf{f}_{\mathbf{p}_n^l}^{(i)}$ is an unbiased estimate of the gradient $ \nabla_{\!\!\mathbf{p}_n^l} \! K_{\!\kappa_n^l}^{\!(i)} ( \mathbf{w}_n^{\!l, (i\!-\!1)}\!\!, \mathbf{p}_{\!n}^l )$ given by $\mathbf{f}_{\mathbf{p}_n^l}^{(i)} \!\!=\!\! (\!1\!\!-\!\!\rho^{(i)}\!)  \mathbf{f}_{\mathbf{p}_n^l}^{(i\!-\!1)} \!\!+
\!\! \frac{\rho^{\!(i)}}{B} \\ \!\!\sum_{\!b=\!1}^B \!\!\nabla_{\!\!\mathbf{p}_n^l} \! g_{\kappa_n^l}^{(i)} ( \mathbf{w}_n^{\!l, (i\!-\!1\!)}\!\!, \mathbf{p}_n^{\!l,(i\!-\!1\!)}\!; \sim \!\! \kappa_n^{\!l,(b)} \!) $ with $\mathbf{f}_{\mathbf{p}_n^l}^{\!(0)} \!\!=\!\! \mathbf{0}$, $\{\rho^{\!(i)}\}$ denoting a sequence of positive numbers that will be specified later, and  $\{\!\sim \!\!\kappa_n^{\!l,(b)}\!\}_{b=1}^B$ representing a mini-batch of $B$ samples generated from the distribution $\prod_{ \bar{\kappa}_n^l \in \sim \kappa_n^{\!l}\!} q(\bar{\kappa}_n^l)$ with the particle weights and positions of $q(\bar{\kappa}_n^l)$ set to those from the latest iteration. Detailed derivation of  $\nabla_{\!\!\mathbf{p}_n^l} \! g_{\kappa_n^l}^{(i)} ( \mathbf{w}_n^{l}, \mathbf{p}_n^{l}; \sim \!\! \kappa_n^{l} )$ is omitted here due to space limitation. Then, by minimizing the surrogate function \eqref{surog fuc, p} subject to constraint \eqref{KLD min subprob p, costraint p},  the intermediate variable $\bar{\mathbf{p}}_n^l = [\bar{p}_{n,1}^l, \dots, \bar{p}_{n,N_p}^l]^T$ can be obtained as $\bar{p}_{n,k}^l =  \dot{\kappa}_n^l - \Delta_n^l $ for $\tilde{p}_{n,k}^l <  \dot{\kappa}_n^l - \Delta_n^l $;  $\bar{p}_{n,k}^l = \tilde{p}_{n,k}^l$ for $\tilde{p}_{n,k}^l$ within the range $[ \dot{\kappa}_n^l - \Delta_n^l,\dot{\kappa}_n^l + \Delta_n^l]$; and $\bar{p}_{n,k}^l =  \dot{\kappa}_n^l + \Delta_n^l $ for $\tilde{p}_{n,k}^l >  \dot{\kappa}_n^l + \Delta_n^l $, where $\tilde{p}_{n,k}^l = p_{n,k}^{l, (i-1)}   - [\mathbf{f}_{\mathbf{p}_n^l}^{(i)}]_k/(2\Omega_{\mathbf{p}_{n}^l} )$, $\forall k$. Finally, $\mathbf{p}_n^{l,(i)}$ is updated by 
\begin{align}
	\label{recur update p}
	\mathbf{p}_n^{l,(i)} = (1-\gamma^{(i)})\mathbf{p}_n^{l,(i-1)} + \gamma^{(i)}\bar{\mathbf{p}}_n^l,
\end{align}
where $\{\gamma^{(i)}\}$ denotes another sequence of positive numbers that will be discussed later.

\par On the other hand, the optimization problem for $\mathbf{w}_n^l$, $\forall \bm{\kappa}^l \in \bm{\kappa}$, $\forall n$, is to minimize $K_{\!\kappa_n^l}^{\!(i)} ( \mathbf{w}_n^{l}, \mathbf{p}_n^{l, (i)} )$ under the constraints \eqref{sum_1 constant, w}. Similarly, we first construct a convex surrogate function for $K_{\!\kappa_n^l}^{\!(i)} ( \mathbf{w}_n^{l}, \mathbf{p}_n^{l, (i)} )$ as follows:
\begin{align}
	\label{surog fuc, w}
	\!\bar{f}_{\!\mathbf{w}_{n}^l}^{(i)} (\mathbf{w}_{\!n}^l ) \!\triangleq\! \big(\! \mathbf{f}_{\mathbf{w}_n^l}^{(i)} \!\big)^{\!\!T}\! \big( \!\mathbf{w}_{n}^l \!\!-\!\! \mathbf{w}_{n}^{l, (i\!-\!1)} \big) \!+\! \Omega_{\mathbf{w}_{n}^l} \| \big( \mathbf{w}_{n}^l \!\!-\!\! \mathbf{w}_{n}^{l, (i\!-\!1)} \|^2, 
\end{align}
where $\Omega_{\mathbf{w}_{n}^l}$ is a positive constant, and $\mathbf{f}_{\mathbf{w}_n^l}^{(i)}$ is an unbiased estimate of the gradient $ \nabla_{\!\!\mathbf{w}_n^l} \!K_{\!\kappa_n^l}^{\!(i)} ( \mathbf{w}_n^{l}, \mathbf{p}_n^{l, (i)} )$ given by $\mathbf{f}_{\mathbf{w}_n^l}^{(i)} \!\!=\!\! (1\!-\!\!\rho^{(i)})  \mathbf{f}_{\mathbf{w}_n^l}^{\!(i\!-\!1)} \!\!+
\! \frac{\rho^{\!(i)}}{B}  \!\!\sum_{\!b=\!1}^B \!\!\nabla_{\!\!\mathbf{w}_n^l} \! g_{\kappa_n^l}^{(i)} ( \mathbf{w}_n^{l, (i\!-\!1\!)}\!, \mathbf{p}_n^{\!l,(i)}; \sim \!\! \kappa_n^{\!l,(b)} ) $ with $\mathbf{f}_{\mathbf{w}_n^l}^{\!(0)} \!\!=\!\! \mathbf{0}$. Detailed derivation of  $\nabla_{\!\!\mathbf{w}_n^l} \! g_{\kappa_n^l}^{(i)} ( \mathbf{w}_n^{l}, \mathbf{p}_n^{\!l}; \sim \!\! \kappa_n^{\!l} )$ is also omitted here. Subsequently, we employ the simplex projection method \cite{chen2011projection} to efficiently minimize the surrogate function \eqref{surog fuc, w} under the constraints \eqref{sum_1 constant, w}, which yields the intermediate variable $\bar{\mathbf{w}}_n^l = [\bar{w}_{n,1}^l, \dots, \bar{w}_{n,N_p}^l]^T$. Finally, $\mathbf{w}_n^{l,(i)}$ is updated by 
\begin{align}
	\label{recur update w}
	\mathbf{w}_n^{l,(i)} = (1-\gamma^{(i)})\mathbf{w}_n^{l,(i-1)} + \gamma^{(i)}\bar{\mathbf{w}}_n^l.
\end{align}
According to \cite[Theorem 3]{hu2024spvbi}, if the sequences $\{\rho^{(i)}\}$ and $\{\gamma^{(i)}\}$ are properly chosen such that the following conditions are satisfied: 
1) $\rho^t \rightarrow 0$, $\sum_t \rho^t = \infty$, $\sum_t (\rho^t)^2 < \infty$; 2) $\lim_{t\rightarrow \infty} \frac{\gamma^t}{\rho^t} = 0$, the SPVBI method almost surely converges to the set of stationary solutions of problem \eqref{KLD min prob}. In practice, we can choose the sequences as $\rho^{(i)} = 5 / (5 + (i-1)^{0.9}, \gamma^{(i)} = 5 / (15 + (i-1))$.

\section*{Appendix C \\  Complexity Analysis}
In this appendix, we analyze the computational complexity of the proposed TS-CET framework in Algorithm 1. Specifically, in stage I, the determination of the coarse 3D set $\mathcal{S}^{(t)}$ has a constant computational cost. Each execution of the tensor-based joint search has a complexity of $\mathcal{O}(R_B \bar{K} L^3)$, while the grid update module requires $\mathcal{O}(N_U^2 + \bar{N}_R^2 + N_f^2)$ for gradient computation. Let $N_{JS}$ denote the average number of the tensor-based joint search executions, then the overall complexity of stage I is $\mathcal{O}\big(N_{\text{I}}(N_{JS} R_B \bar{K} L^3 + N_U^2 + \bar{N}_R^2 + N_f^2)\big)$.
In stage II, the SPVBI method in Module A adopts mini-batch sampling and surrogate function minimization, resulting in a per-iteration complexity of $\mathcal{O}\big(2J(N_p B F_{\text{grad}} + N_p^3)\big)$, where $F_{\text{grad}}$ is the average number of floating point operations (FLOPs) required for a single gradient update, $N_p^3$ represents the complexity of minimizing surrogate function, and $J$ is the number of variables in $\boldsymbol{\kappa}$. Message passing in Module B has linear complexity. Let $N_{SSCA}$ denote the average number of SSCA iterations; then the total complexity of stage II is $\mathcal{O}\big(2J N_{\text{II}} N_{SSCA}(N_p B F_{\text{grad}} + N_p^3)\big)$.
Therefore, the overall computational complexity of TS-CET is $\mathcal{O}\big(N_{\text{I}}(N_{JS} R_B \bar{K} L^3 + N_U^2 + \bar{N}_R^2 + N_f^2) + 2J N_{\text{II}} N_{SSCA}(N_p B F_{\text{grad}} + N_p^3)\big)$.

\bibliographystyle{IEEEtran}
\bibliography{bib_tracking_IRS}

@ARTICLE{cui2022nearfield,
	author={Cui, Mingyao and Dai, Linglong},
	journal={IEEE Trans. Commun.}, 
	title={Channel Estimation for Extremely Large-Scale {MIMO}: Far-Field or Near-Field?}, 
	year={2022},
	volume={70},
	number={4},
	pages={2663-2677},
	  month={Apr.},
	doi={10.1109/TCOMM.2022.3146400}}

@article{kolda2009tensor,
	title={Tensor decompositions and applications},
	author={Kolda, Tamara G and Bader, Brett W},
	journal={SIAM Rev.},
	volume={51},
	number={3},
	pages={455--500},
	year={2009},
	month={Aug.},
	publisher={SIAM}
}

@ARTICLE{yu2023mixednf,
	author={Lu, Yu and Dai, Linglong},
	journal={IEEE Trans. Commun.}, 
	title={Near-Field Channel Estimation in Mixed {LoS}/{NLoS} Environments for Extremely Large-Scale {MIMO} Systems}, 
	year={2023},
	volume={71},
	number={6},
	pages={3694-3707},
	doi={10.1109/TCOMM.2023.3260242},
	ISSN={1558-0857},
	month={Jun.},}

@ARTICLE{Selvan2017fraun,
	author={Selvan, Krishnasamy T. and Janaswamy, Ramakrishna},
	journal={IEEE Antennas Propag. Mag.}, 
	title={Fraunhofer and {Fresnel} Distances: Unified derivation for aperture antennas}, 
	year={2017},
	volume={59},
	number={4},
	pages={12-15},
	doi={10.1109/MAP.2017.2706648},
	ISSN={1558-4143},
	month={Aug.},}

@ARTICLE{yang2024nearIRS,
  author  = {S. Yang and C. Xie and W. Lyu and B. Ning and Z. Zhang and C. Yuen},
  title   = {Near-Field Channel Estimation for Extremely Large-Scale Reconfigurable Intelligent Surface ({XL-RIS})-Aided Wideband {mmWave} Systems},
  journal = {IEEE J. Sel. Areas Commun.},
  volume  = {42},
  number  = {6},
  pages   = {1567--1582},
  year    = {2024},
  month   = {Jun.},
  doi     = {10.1109/JSAC.2024.3389120}
}

@ARTICLE{pan2023irsloc,
	author={Pan, Yijin and Pan, Cunhua and Jin, Shi and Wang, Jiangzhou},
	journal={IEEE J. Sel. Top. Signal Process.}, 
	title={{RIS}-Aided Near-Field Localization and Channel Estimation for the Terahertz System}, 
	year={2023},
	volume={17},
	number={4},
	pages={878-892},
	doi={10.1109/JSTSP.2023.3285431},
	ISSN={1941-0484},
	month={Jul.},}

@ARTICLE{3GPPzc,
	author={},
	journal={3GPP, TS 38.211 V16.7.0 Release 16, Technical Specification Group Radio Access Network; NR; Physical channels and modulation}, 
	title={}, 
	year={2021},
	volume={},
	number={},
	pages={},
	doi={10.1109/JSTSP.2023.3285431},
	ISSN={1941-0484},
	month={Sep.},}

@ARTICLE{zhou2022irsce,
	author={Zhou, Gui and Pan, Cunhua and Ren, Hong and Popovski, Petar and Swindlehurst, A. Lee},
	journal={IEEE Trans. Signal Process.}, 
	title={Channel Estimation for {RIS}-Aided Multiuser Millimeter-Wave Systems}, 
	year={2022},
	volume={70},
	number={},
	pages={1478-1492},
	month={Mar.},
	doi={10.1109/TSP.2022.3158024}}

@ARTICLE{wang2021joint,
		author={Wang, Peilan and Fang, Jun and Dai, Linglong and Li, Hongbin},
		journal={IEEE Trans. Wireless Commun.}, 
		title={Joint Transceiver and Large Intelligent Surface Design for Massive {MIMO} {mmWave} Systems}, 
		year={2021},
		volume={20},
		number={2},
		pages={1052-1064},
		doi={10.1109/TWC.2020.3030570},
		ISSN={1558-2248},
		month={Feb.},}

@ARTICLE{Donoughue2012CNN,
	author={O'Donoughue, Nicholas and Moura, José M. F.},
	journal={IEEE Trans. Signal Process.}, 
	title={On the Product of Independent Complex {Gaussians}}, 
	year={2012},
	volume={60},
	number={3},
	pages={1050-1063},
	doi={10.1109/TSP.2011.2177264},
	ISSN={1941-0476},
	month={Mar.},}

@INPROCEEDINGS{som2011approximate,
	author={Som, Subhojit and Schniter, Philip},
	booktitle={Proc. Int. Conf. Mach. Learn. (ICML)}, 
	title={Approximate message passing for recovery of sparse signals with Markov-random-field support structure}, 
	year={2011},
	volume={},
	number={},
	pages={},
	doi={},
	ISSN={},
	month={Jul.},}

@ARTICLE{kuai2019struct,
	author={Kuai, Xiaoyan and Chen, Lei and Yuan, Xiaojun and Liu, An},
	journal={IEEE Trans. Wireless Commun.}, 
	title={Structured Turbo Compressed Sensing for Downlink Massive {MIMO-OFDM} Channel Estimation}, 
	year={2019},
	volume={18},
	number={8},
	pages={3813-3826},
	doi={10.1109/TWC.2019.2917905},
	ISSN={1558-2248},
	month={Aug.},}

@ARTICLE{xu2023successive,
  author={Xu, Wenkang and Liu, An and Zhou, Bingpeng and Zhao, Min-Jian},
  journal={IEEE Transactions on Wireless Communications}, 
  title={Successive Linear Approximation {VBI} for Joint Sparse Signal Recovery and Dynamic Grid Parameters Estimation}, 
  year={2025},
  month={Nov.},
  volume={24},
  number={11},
  pages={9645-9659},
  doi={10.1109/TWC.2025.3574489}}

@ARTICLE{ziniel2012dynamic,
	author={Ziniel, Justin and Schniter, Philip},
	journal={IEEE Trans. Signal Process.}, 
	title={Dynamic Compressive Sensing of Time-Varying Signals Via Approximate Message Passing}, 
	year={2013},
	volume={61},
	number={21},
	pages={5270-5284},
	doi={10.1109/TSP.2013.2273196},
	ISSN={1941-0476},
	month={Nov.},}

@INPROCEEDINGS{Zhang2016tracking,
	author={Zhang, Chuang and Guo, Dongning and Fan, Pingyi},
	booktitle={Proc. IEEE Int. Conf. Commun. (ICC)}, 
	title={Tracking angles of departure and arrival in a mobile millimeter wave channel}, 
	year={2016},
	volume={},
	number={},
	pages={1-6},
	doi={10.1109/ICC.2016.7510902},
	ISSN={1938-1883},
	month={May},}

@INPROCEEDINGS{Cao2017semidefinite,
  author={Cao, Pan and Thompson, John and Poor, H. Vincent},
  booktitle={2017 25th European Signal Processing Conference (EUSIPCO)}, 
  title={A sequential constraint relaxation algorithm for rank-one constrained problems}, 
  year={2017},
  volume={},
  number={},
  pages={1060-1064},
  doi={10.23919/EUSIPCO.2017.8081370},
  ISSN={2076-1465},
  month={Aug},}

@ARTICLE{liu2020tvbi,
	author={Liu, An and Liu, Guanying and Lian, Lixiang and Lau, Vincent K. N. and Zhao, Min-Jian},
	journal={IEEE Trans. Wireless Commun.}, 
	title={Robust Recovery of Structured Sparse Signals With Uncertain Sensing Matrix: A Turbo-{VBI} Approach}, 
	year={2020},
	volume={19},
	number={5},
	pages={3185-3198},
	doi={10.1109/TWC.2020.2971193},
	ISSN={1558-2248},
	month={May},}

@ARTICLE{dai2018sbl,
	author={Dai, Jisheng and Liu, An and Lau, Vincent K. N.},
	journal={IEEE Trans. Signal Process.}, 
	title={{FDD} Massive {MIMO} Channel Estimation With Arbitrary {2D}-Array Geometry}, 
	year={2018},
	volume={66},
	number={10},
	pages={2584-2599},
	doi={10.1109/TSP.2018.2807390},
	ISSN={1941-0476},
	month={May},}

@ARTICLE{zhou2017pvbi,
	author={Zhou, Bingpeng and Chen, Qingchun and Wymeersch, Henk and Xiao, Pei and Zhao, Lian},
	journal={IEEE Trans. Mobile Comput.}, 
	title={Variational Inference-Based Positioning with Nondeterministic Measurement Accuracies and Reference Location Errors}, 
	year={2017},
	volume={16},
	number={10},
	pages={2955-2969},
	doi={10.1109/TMC.2016.2640294},
	ISSN={1558-0660},
	month={Oct.},}

@book{parisi1988statistical,
	title={Statistical field theory},
	author={Parisi, Giorgio and Shankar, Ramamurti},
	year={1988},
	publisher={Westview Press}
}

@ARTICLE{tan2018omp,
	author={Tan, Guoping and Wu, Bingyang and Herfet, Thorsten},
	journal={IEEE Trans. Wireless Commun.}, 
	title={Performance Analysis of {OMP}-Based Channel Estimations in Mobile {OFDM} Systems}, 
	year={2018},
	volume={17},
	number={5},
	pages={3459-3473},
	doi={10.1109/TWC.2018.2813380},
	ISSN={1558-2248},
	month={May},}

@article{luo2010semidefinite,
	title={Semidefinite relaxation of quadratic optimization problems},
	author={Luo, Zhi-Quan and Ma, Wing-Kin and So, Anthony Man-Cho and Ye, Yinyu and Zhang, Shuzhong},
	journal={IEEE Signal Process. Mag.},
	volume={27},
	number={3},
	pages={20--34},
	year={2010},
	month={May},
}

@ARTICLE{liu2019ssca,
	author={Liu, An and Lau, Vincent K. N. and Kananian, Borna},
	journal={IEEE Trans. Signal Process.}, 
	title={Stochastic Successive Convex Approximation for Non-Convex Constrained Stochastic Optimization}, 
	year={2019},
	volume={67},
	number={16},
	pages={4189-4203},
	doi={10.1109/TSP.2019.2925601},
	ISSN={1941-0476},
	month={Aug.},}

@article{chen2011projection,
	title={Projection onto a simplex},
	author={Chen, Yunmei and Ye, Xiaojing},
	journal={arXiv preprint arXiv:1101.6081},
	year={2011}
}

@ARTICLE{hu2024spvbi,
	author={Hu, Zhixiang and Liu, An and Wan, Yubo and Han, Tony Xiao and Zhao, Minjian},
	journal={IEEE Internet Things J.}, 
	title={A Two-Stage Multiband Delay Estimation Scheme via Stochastic Particle-Based Variational Bayesian Inference}, 
	year={2024},
	volume={11},
	number={11},
	pages={19632-19645},
	doi={10.1109/JIOT.2024.3367752},
	ISSN={2327-4662},
	month={Jun.},}

@book{li2009markov,
	title={Markov random field modeling in image analysis},
	author={Li, Stan Z},
	year={2009},
	publisher={Springer Science \& Business Media}
}

@ARTICLE{chen2023path,
  author={Chen, Zijian and Zhao, Ming-Min and Li, Min and Lei, Ming and Zhao, Min-Jian},
  journal={IEEE Trans. Wireless Commun.},
  title={{IRS}-Aided Joint Spatial Division and Multiplexing for {mmWave} Multiuser {MISO} Systems},
  year={2023},
  volume={22},
  number={11},
  pages={7789--7804},
  doi={10.1109/TWC.2023.3255551},
  month={Nov.},
  issn={1558-2248},
}

@ARTICLE{Wu2023parametric,
  author={Wu, Jiao and Kim, Seungnyun and Shim, Byonghyo},
  journal={IEEE Trans. Commun.}, 
  title={Parametric Sparse Channel Estimation for {RIS}-Assisted Terahertz Systems}, 
  year={2023},
  month={Sep.},
  volume={71},
  number={9},
  pages={5503-5518},
  doi={10.1109/TCOMM.2023.3285759}}

@ARTICLE{Daniel2019tensor,

  author={Araújo, Daniel Costa and de Almeida, André L. F. and Da Costa, João P. C. L. and de Sousa, Rafael T.},
  journal={IEEE Access}, 
  title={Tensor-Based Channel Estimation for Massive {MIMO}-{OFDM} Systems}, 
  year={2019},
  volume={7},
  number={},
  pages={42133-42147},
  doi={10.1109/ACCESS.2019.2908207}}

@article{Beck2010SCA,
  title={A sequential parametric convex approximation method with applications to nonconvex truss topology design problems},
  author={ Beck, Amir  and  Ben-Tal, Aharon  and  Tetruashvili, Luba },
  journal={J. Global Optim.},
  volume={47},
  number={1},
  pages={29-51},
  year={2010},
}

@ARTICLE{Wu2021IRS,
  author={Wu, Qingqing and Zhang, Shuowen and Zheng, Beixiong and You, Changsheng and Zhang, Rui},
  journal={IEEE Trans. Commun.}, 
  title={Intelligent Reflecting Surface-Aided Wireless Communications: A Tutorial}, 
  year={2021},
  volume={69},
  number={5},
  month={May},
  pages={3313-3351},
  doi={10.1109/TCOMM.2021.3051897}}

@ARTICLE{Wei2022narrow,

  author={Wei, Yi and Zhao, Ming-Min and Zhao, Min-Jian and Cai, Yunlong},

  journal={IEEE Trans. Wireless Commun.}, 

  title={Channel Estimation for {IRS}-Aided Multiuser Communications With Reduced Error Propagation}, 

  year={2022},
  month ={Apr.},
  volume={21},

  number={4},
    
  pages={2725-2741},

  doi={10.1109/TWC.2021.3115161}}

@ARTICLE{saa20206G,
  author={Saad, Walid and Bennis, Mehdi and Chen, Mingzhe},
  journal={IEEE Network}, 
  title={A Vision of 6{G} Wireless Systems: Applications, Trends, Technologies, and Open Research Problems}, 
  year={2020},
  volume={34},
  number={3},
  pages={134-142},
  doi={10.1109/MNET.001.1900287}}

@ARTICLE{Lu2024NF,
  author={Lu, Haiquan and Zeng, Yong and You, Changsheng and Han, Yu and Zhang, Jiayi and Wang, Zhe and Dong, Zhenjun and Jin, Shi and Wang, Cheng-Xiang and Jiang, Tao and You, Xiaohu and Zhang, Rui},
  journal={IEEE Commun. Surv. Tutor.}, 
  title={A Tutorial on Near-Field {XL-MIMO} Communications Toward 6{G}}, 
  year={2024},
  volume={26},
  number={4},
  pages={2213-2257},
  doi={10.1109/COMST.2024.3387749}}

@ARTICLE{Zheng2022CSCE,
  author={Zheng, Xi and Wang, Peilan and Fang, Jun and Li, Hongbin},
  journal={IEEE Wireless Commun. Lett.}, 
  title={Compressed Channel Estimation for {IRS}-Assisted Millimeter Wave {OFDM} Systems: A Low-Rank Tensor Decomposition-Based Approach}, 
  year={2022},
  month={Jun.},
  volume={11},
  number={6},
  pages={1258-1262},
  doi={10.1109/LWC.2022.3163661}}

@ARTICLE{you20256G,
  author={You, Changsheng and Cai, Yunlong and Liu, Yuanwei and Di Renzo, Marco and Duman, Tolga M. and Yener, Aylin and Lee Swindlehurst, A.},
  journal={IEEE J. Sel. Areas Commun.}, 
  title={Next Generation Advanced Transceiver Technologies for 6{G} and Beyond}, 
  year={2025},
  month={Mar.},
  volume={43},
  number={3},
  pages={582-627},
  doi={10.1109/JSAC.2025.3536557}}
\end{document}